\documentclass[twocolappendix]{emulateapj}
\usepackage{color}

\begin{document}

\title{Searching for light echoes due to CSM in SN Ia spectra}
\author{Sebasti\'an Marino$^{2,1}$, Santiago Gonz\'alez-Gait\'an$^{2,1}$, Francisco F\"orster$^{2,3}$, Gast\'on Folatelli$^{4,5}$, Mario Hamuy$^{1,2}$ \& Eric Hsiao$^{6,7}$}
\affiliation{$^{1}$Departamento de Astronom\'ia, Universidad de Chile, 1515 Camino El Observatorio, Las Condes, Santiago, Chile}
\affiliation{$^2$Millennium Institute of Astrophysics, Casilla 36-D, Santiago, Chile}
\affiliation{$^{3}$Centro de Modelamiento Matem\'atico, Universidad de Chile, Av. Blanco Encalada 2120 Piso 7, Santiago, Chile}
\affiliation{$^{4}$Instituto de Astrof\'isica de La Plata (IALP, CONICET), Argentina.}
\affiliation{$^{5}$Kavli Institute for the Physics and Mathematics of the Universe, the University of Tokyo, Kashiwa, Japan 277-8583 (Kavli IPMU, WPI)}
\affiliation{$^{6}$Carnegie Observatories, Las Campanas Observatory, Casilla 601, La Serena, Chile}
\affiliation{$^{7}$Department of Physics and Astronomy, Aarhus University, Ny Munkegade 120, DK-8000 Aarhus C, Denmark}

\begin{abstract}
We present an analytical model for light echoes (LEs) coming from
circumstellar material (CSM) around Type Ia Supernovae (SNe Ia). Using
this model we find two spectral signatures at 4100 \AA \ and 6200 \AA
\ that are useful to identify LEs during the Lira law phase (between
35 and 80 days after maximum light) coming from nearby CSM at
distances of 0.01-0.25 pc. We analyze a sample of 89 SNe Ia divided in
two groups according to their $B-V$ decline rate during the Lira law
phase, and search for LEs from CSM interaction in the group of SNe
with steeper slopes by comparing their spectra with our LE model. We
find that a model with LEs + pure extinction from interstellar
material (ISM) fits better the observed spectra than a pure ISM
extinction model that is constant in time, but we find that a
decreasing extinction alone explains better the observations without
the need of LEs, possibly implying dust sublimation due to the
radiation from the SN.

\end{abstract}

\section{Introduction}

Type Ia supernovae (SNe~Ia) are one of the most studied objects in
astronomy. Owing to their standardizable high luminosities that make
them unrivaled distance indicators up to high redshifts ($z\sim2$,
\citealt{Jones13}), astronomers searching for the ultimate fate of the
universe have studied them in ever greater detail, discovering several
thousands of them up to now \citep[e.~g.][]{Sako14}. Despite the
success as cosmological probes, increasing observations and clues
reveal that the puzzle of the nature and mechanism generating these
colossal explosions is still far from reaching a conclusive solution.

As best shown in recent well-studied nearby objects
\citep{Nugent11,Bloom12}, SNe~Ia may originate from the explosion of a
compact CO-rich white dwarf (WD) in a binary system. A common
candidate considered for the binary companion has for long been a
non-degenerate star such as a main-sequence or a red giant star, that
donates mass to the WD (single degenerate, SD, scenario) either in a
stable fashion so that the WD nears the Chandrasekhar mass and
explodes first subsonically and then supersonically \citep[Ch-SD,
  e.~g.][]{Blondin13,Sim13,Roepke12} or via unstable accretion leading
to an initial detonation in the outer layer of the WD that triggers a
subsequent detonation near the center prior to reaching the
Chandrasekhar mass \citep[sub Ch-SD, e.~g.][]{Sim12,Kromer10}.

From an observational point of view, the SD scenario model has both
evidence for and against. Among the observations that disfavor the
model are: the absence of hydrogen and helium in their spectra
\citep{Lundqvist13,Shappee13}; as well as the absence of radio and
X-ray emission \citep{Horesh12,Chomiuk12} which set tight constraints
on mass loss from a progenitor; the non-detection of early emission
from shock interaction with a companion
\citep{Bloom12,Bianco11,Hayden10b} and the pre-explosion
non-detections \citep{Li11} that generally rule out red giants and
He-stars; the lack of sufficient galactic X-ray emission
\citep{Gilfanov10,DiStefano10} and UV radiation
\citep{Woods13,Johansson14} expected from mass accretion in the Ch-SD
scenario; and the measured SN Ia rate as a function of redshift which
challenges the modelled delay time distribution for the classical
Ch-SD scenario \citep[][and references therein]{Maoz12rev}. Some of
these issues can be addressed through different alternative SD models
\citep{Justham11,DiStefano11}.

Among the observational evidence that favors the SD channel is the
presence of nearby circumstellar material (CSM), presumably from mass
loss in the progenitor system prior to explosion, manifested from:
CSM/ejecta interaction \citep{Hamuy03b,Silverman13csm}, and most
notably in the case of PTF11kx \citep{Dilday12}; the discovery of
narrow Na I D absorption lines that vary with time
\citep{Patat07,Blondin09,Simon09,Sternberg13}; and the statistical
preference for the interstellar lines to show blueshifts
\citep{Sternberg11,Maguire13,Phillips13}. Altogether, the
observational evidence suggests the possibility of multiple channels
for SNe Ia.

Additionally, it has been suggested that such nearby CSM could affect
the colors of SNe~Ia through light scattering in the line of sight and
explain in this way some of the differences in total to selective
extinction ratios ($R_V$) found in SN hosts compared to the Milky Way,
MW \citep{Wang05,Goobar08,Amanullah11}. Heavily extincted SNe clearly
show a different $R_V$ while SNe with moderate extinction show values
consistent with the MW (e.g. \citealt{Mandel11},
\citealt{Burns14}). Cosmological studies using standard light-curve
fitters obtain a luminosity-color relation that suggest reddening laws
lower than the MW \citep[e.g][]{Guy05}, however it is possible that SN
intrinsic colors are more complicated and incorrectly modeled
\citep[e.g.][]{Scolnic2014,Chotard11,Conley07}.

Understanding the origin of the dispersion of SNe Ia colors not only
affords the opportunity to understand their nature, but also to remove
a major source of systematic uncertainty in SN Ia cosmology. As shown
in \citet{Kim13}, color might indeed be the principal parameter of
diversity in SN Ia light-curves, followed only by the well-known
light-curve width parameter of SNe~Ia. \citet[][hereafter
  F13]{Forster13} showed that the SN Ia color evolution with time is
related to the strength of the narrow sodium absorption, suggesting
that at least some part of it might originate from a closer
interaction with dust than with host interstellar medium (ISM). In
particular, they found that redder objects at maximum light have
stronger narrow absorption lines and evolve faster from red to blue
during the late time evolution of the Lira law decline of 30 to 90
days past maximum light \citep{Lira95,Phillips99}. Possible explanations for
this are light-echoes from CSM that affect the late-time colors or,
alternatively, CSM dust sublimation in the line of sight.

In this paper, we aim to investigate the results of F13 further and
test the hypothesis of nearby CSM by looking for spectroscopic
signatures of light-echoes in a large sample of nearby SN Ia spectra
in the Lira law phase. Echoes from nearby ISM (and possible CSM) have
previously been reported for SN1991T, SN1998bu, SN2006X, SN1992G and
SN2014J at nebular phases
\citep{Schmidt94c,Cappellaro01,Wang08b,Silverman13neb,crotts2014}. These
individual studies focused on echoes generated at large distances from
the SN, tens to hundreds of parsecs away, scattering hundreds of days
past maximum light.

Here we search for light echoes (LEs) at earlier times ($>30d$ past
maximum light), coming from nearby CSM dust that is at less than a
parsec from the SN. Such CSM can potentially affect the colors and the
Lira law decline rate \citep{Amanullah11}. Hence we search for LEs in
the group of SNe analyzed in F13 that presented more extinction and a
steeper than normal $B-V$ evolution (hereafter fast Lira decliners) to
test the hypothesis that these may originate in regions of nearby
CSM. To do this, we use SNe with a shallow Lira law slope (hereafter
slow Lira decliners) as a reference set of SNe without CSM
interaction.

In \S\ref{model} we present our simple LE model. Then in
\S\ref{predictions} we focus on the prediction of observable
spectroscopy features to look for light echoes. In \S\ref{analysis} we
present the data and the analysis. \S\ref{results} summarizes the
results of our search for LEs and in \S\ref{discussion} we discuss the
success of our LE model, the validity of our assumptions and other
possible mechanisms that could explain fast Lira decliners. Finally,
the main conclusions are summarized in \S\ref{conclusions}.

\section{Light echo model}
\label{model}

The effects of the interaction of light with intervening dust from CSM
causing scattering away from the line of sight, and therefore
extinction and reddening, has been studied and modeled in depth in the
past \citep[e.~g.][]{Chevalier86,Wang05,Patat05,Patat06,Goobar08}. We
present here a simple analytical model that is easy to implement
numerically and makes clear observable predictions to directly compare
with data. The CSM consists of a simple spherically isotropic shell of
dust (with $R_{V}=3.1$), that absorbs and scatters the light of the
SN. The radius of the shell is initially fixed at 0.05 pc to produce
LEs reaching the observer with time delays of $\sim$50 days and
affecting the colors during the Lira law phase, as light emitted at
maximum light is observed at later epochs. At these distances we
expect the temperature of the dust to be slightly lower than the
sublimation temperature (2,000 K).  We only consider single scattering
for simplicity and because multiple scattering in the CSM becomes
important when its optical depth is larger than 1
\citep[see][]{Patat05}. According to our analysis all the SNe we
considered have a total optical depth $\lesssim 1$ in the visible,
with the exception of SN1997cw, SN1999gd, SN2003cg and SN2006X which
have a total extinction $A_{V}>1$ (see \S\ref{extinctionlaw}); but
their $A_{V}$ due to the CSM extinction is lower than unity according
to our models (see \S \ref{Res: LE}). Hence, ignoring multiple
scattering is a reasonable approximation.

\begin{figure}[h!]
\centering
\includegraphics[trim=0.0cm 7.5cm 5.0cm 0.0cm, clip=true,width=0.8\linewidth]{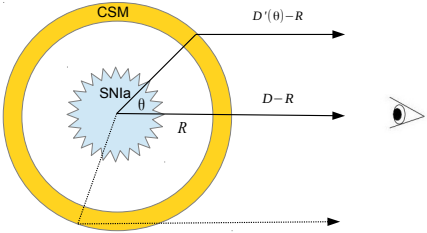}
\caption{Sketch of the CSM model. Part of the light emitted by the SN is scattered by the CSM and redirected to the observer arriving as a light echo delay.}\label{illustration}
\end{figure}

We assume that the observed flux is the sum of the light coming
directly from the SN and the SN light scattered by the CSM, i.e. light
echoes. The direct flux contribution from the SN at a given epoch $t$
that is extincted and scattered by intervening dust \emph{without}
including the contribution from LEs, can be written as:

\begin{equation}
f(t,\lambda)=f^{0}(t,\lambda)e^{-[\sigma_{s}(\lambda)+\sigma_{a}(\lambda)]N_{\mathrm{ tot}}} \label{eq:pureext}
\end{equation}  
where $f^{0}$ is the intrinsic flux of the SN, $\sigma_{s}(\lambda)$
and $\sigma_{a}(\lambda)$ are the cross sections of the dust particles
for scattering and absorption of photons at wavelength $\lambda$,
respectively. $N_{\mathrm{tot}}$ is the total column density of dust
between the SN and the observer and is equal to
$N_{\mathrm{ISM}}+N_{\mathrm{CSM}}$. Then, the total observed flux
adding the LE contribution can be expressed as:
\begin{eqnarray}
F(t,\lambda)&=&f(t,\lambda)+\mathrm{LE} \nonumber \\
&=&f^{0}(t,\lambda)e^{-\sigma(\lambda)N_{\mathrm{tot}}}+S(t,\lambda)e^{-\sigma(\lambda)N_{\mathrm{ISM}}} \label{eq:observed spec}
\end{eqnarray}
where the first term is the intrinsic flux of the SN extincted by
the total column density of dust (eq. \ref{eq:pureext}) and the second
term is the contribution from LEs, $S(t,\lambda)$, extincted by
ISM dust. $\sigma(\lambda)$ is the sum of the scattering and
absorption cross section. We can express $S(t,\lambda)$ as the sum of
the light scattered by the CSM at different angles and epochs:
\begin{eqnarray}
S(t,\lambda)&=&\int f^{0}(tr,\lambda)w(\lambda)(1-e^{-\sigma(\lambda)N_{\mathrm{CSM}}})\Phi(\theta,\lambda) d\Omega \hspace{0.5cm} \\
tr&=&t-\frac{D'(\theta)-D}{c} \approx t-\frac{R(1-cos(\theta))}{c} \label{tr}
\end{eqnarray} 
where $w(\lambda)(1-e^{-\sigma(\lambda)N_{\mathrm{CSM}}})$ represents
the fraction of light scattered by the CSM, which has a column density
of $N_{\mathrm{CSM}}$. $c$ is the speed of light, $w(\lambda)$ is
the dust albedo and $tr$ is a pseudo retarded time, i.e. the time at
which a scattered pulse of light has to be emitted to reach the
observer at the same time than a pulse emitted at time $t$ going
straight to the observer. $D$ is the distance between the photosphere
of the SN and the observer and $D'(\theta)$ is the path length
travelled by a photon being scattered by the CSM to the observer at an
angle $\theta$ (see figure \ref{illustration}).
 
$\Phi(\theta,\lambda)$ is the scattering phase function proposed in \citet{Henyey41}:
\begin{equation}
\Phi(\theta,\lambda)=\frac{1}{4\pi}\frac{1-g(\lambda)^{2}}{(1+g(\lambda)^{2}-2g(\lambda)cos(\theta))^{3/2}} \label{phi}
\end{equation}
where $g$ is the degree of forward scattering. When $g=1$ we have
complete forward scattering and $g=0$ means isotropic scattering
($\Phi(\theta,\lambda)=1/(4\pi) $). We also define the parameter
$f_{\mathrm{CSM}}\equiv N_{\mathrm{CSM}}/N_{\mathrm{tot}}$ to simplify
the notation, taking values between 0 and 1. Finally, defining a delay
parameter $\tau\equiv t-tr$, an extinction factor $X(\lambda)\equiv
e^{-(\sigma_{s}(\lambda)+\sigma_{a}(\lambda))N_{\mathrm{tot}}}$ and
making a change of variable, using equation \ref{tr}, we obtain
\begin{equation}
\mbox{\small $S(t,\lambda)=\frac{\omega(\lambda)(1-X^{f_{\mathrm{CSM}}} )}{\tau_{\max}}\int^{\tau_{\max}}_{0} f^{0}(t-\tau,\lambda)\Phi'(\tau,\lambda)d\tau$ }  \label{eq: echo}
\end{equation} 
\noindent where $\tau_{\max}=2R/c$ is the maximum delay for a light
echo ($\theta=\pi$) and
$\Phi'(\tau,\lambda)=4\pi\Phi(\theta(\tau),\lambda)$. We performed
this integral numerically using the Simpson's 1/3rd integration rule
and a time step of one day to simulate spectra and to fit this model
to real data.

\section{Light echo model predictions}\label{predictions}

\begin{figure}[h!]
\centering
\includegraphics[width=1.0\linewidth]{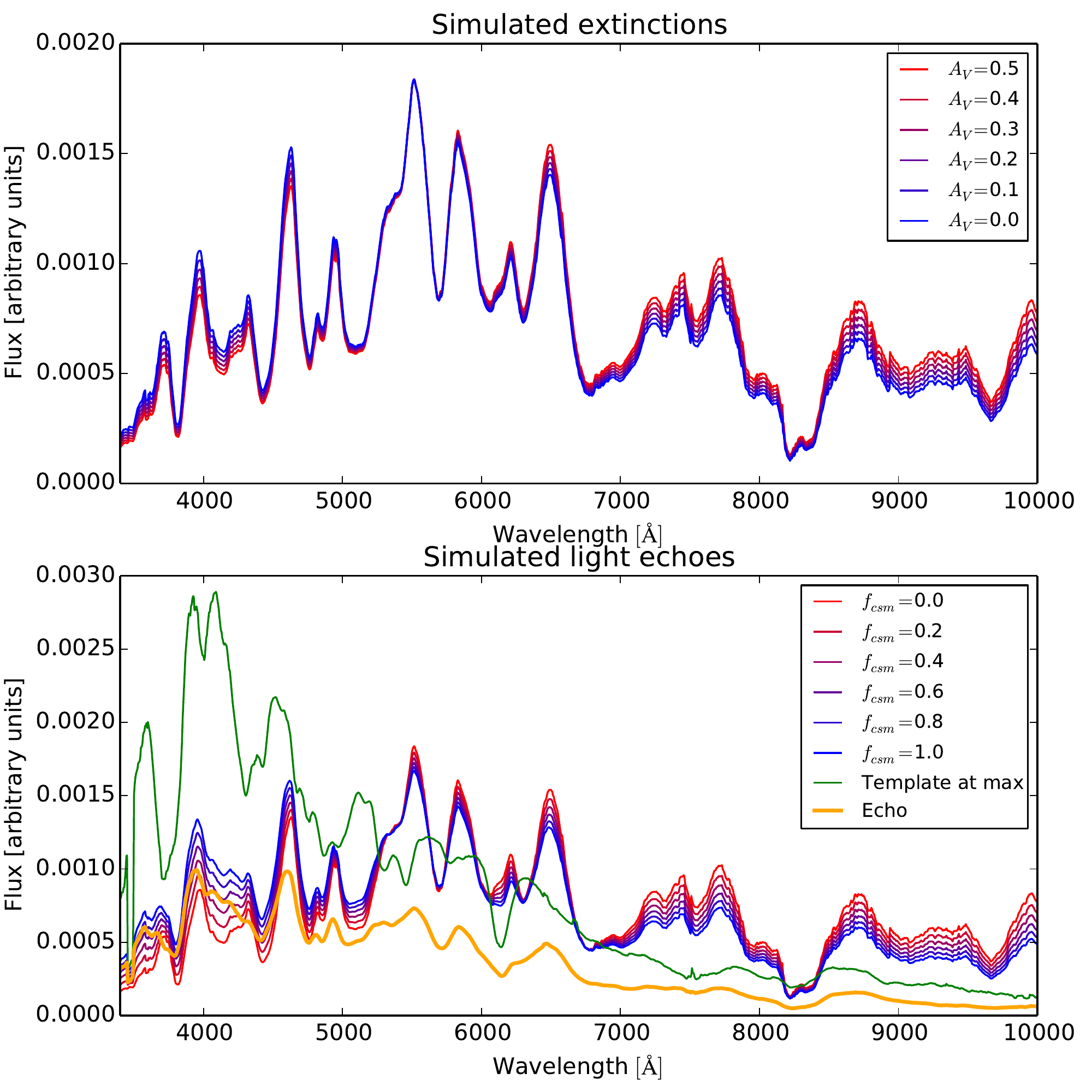}
\caption{Upper panel: simulated spectra at 50 days past maximum
  light extincted by different amounts of dust ($A_V$=0-0.5) with
  the same $R_V$=3.1 extinction law. All spectra have been normalized
  to the same V-band flux. Bottom panel: simulated spectra at 50
  days past maximum light with the same reddening law, but also with
  LEs due to CSM are shown. We fixed $A_{V}=0.5$ and varied the
  fractional amount of CSM ($f_{\mathrm{CSM}}$) with a radius of 0.05
  pc. The green line represents a typical maximum
    light spectrum, while in orange the LE spectrum is shown on an
    arbitrary scale. All models are normalized to the same $V$-band
  flux.}\label{echovsext}
\end{figure}

To simulate spectra with different extinctions and LEs using
eq. \ref{eq: echo} we need to adopt a dust albedo, an extinction law,
a phase function or $g(\lambda)$, and spectral templates with no
extinction nor LEs at different epochs. For the extinction law we take
the parametrization proposed in \citet{Fitzpatrick99}. We use the
albedo $w(\lambda)$ and the degree of forward scattering $g(\lambda)$
from the MW used in \citet{Goobar08} which accounts for the dust
properties of the CSM. We construct unreddened spectral templates at
different epochs from weighted bootstrapped averages of observed
spectra of slow $B-V$ Lira law decliners (see section
\ref{templates}), together with light-curve templates that we need
since we normalize the spectra by their $V$-band flux (see
section \ref{data}). In Figure \ref{echovsext} we show the different
scenarios for late-time (Lira law phase) model spectra when pure
extinction and simulated LEs affect the SN emission.

We search for a way to distinguish if part of the dust found at
maximum light is producing light echoes. LE spectra are integrated
spectra weighted by the light-curve, and thus dominated by spectra
around peak (see Figure \ref{echovsext} with peak template spectrum
and LE spectrum). LE spectra are blue and have very strong broad
emission and absorption lines, with prominent peaks at 4000, 4600,
4900 \AA \ and minima at 4400 and 6200 \AA.

When LE spectra are added to SN spectra: (1) the fact that the LE
spectra are blue has a low-order effect on the observed spectra by
making the colors bluer, similar to less reddening and thus difficult
to differentiate; (2) the strong broad lines add an additional
modulation to the observed spectra that is very distinct to the effect
of reddening, since it introduces differences on scales of a couple of
hundred armstrongs. By looking specifically at the wavelengths where
the LE spectra has peaks or minima, it is possible to differentiate
between the two scenarios. In \S~\ref{analysis} we compare these
simulations to the observed spectra of SNe~Ia.

\begin{figure}[h!]
\centering
\includegraphics[width=1.0\linewidth]{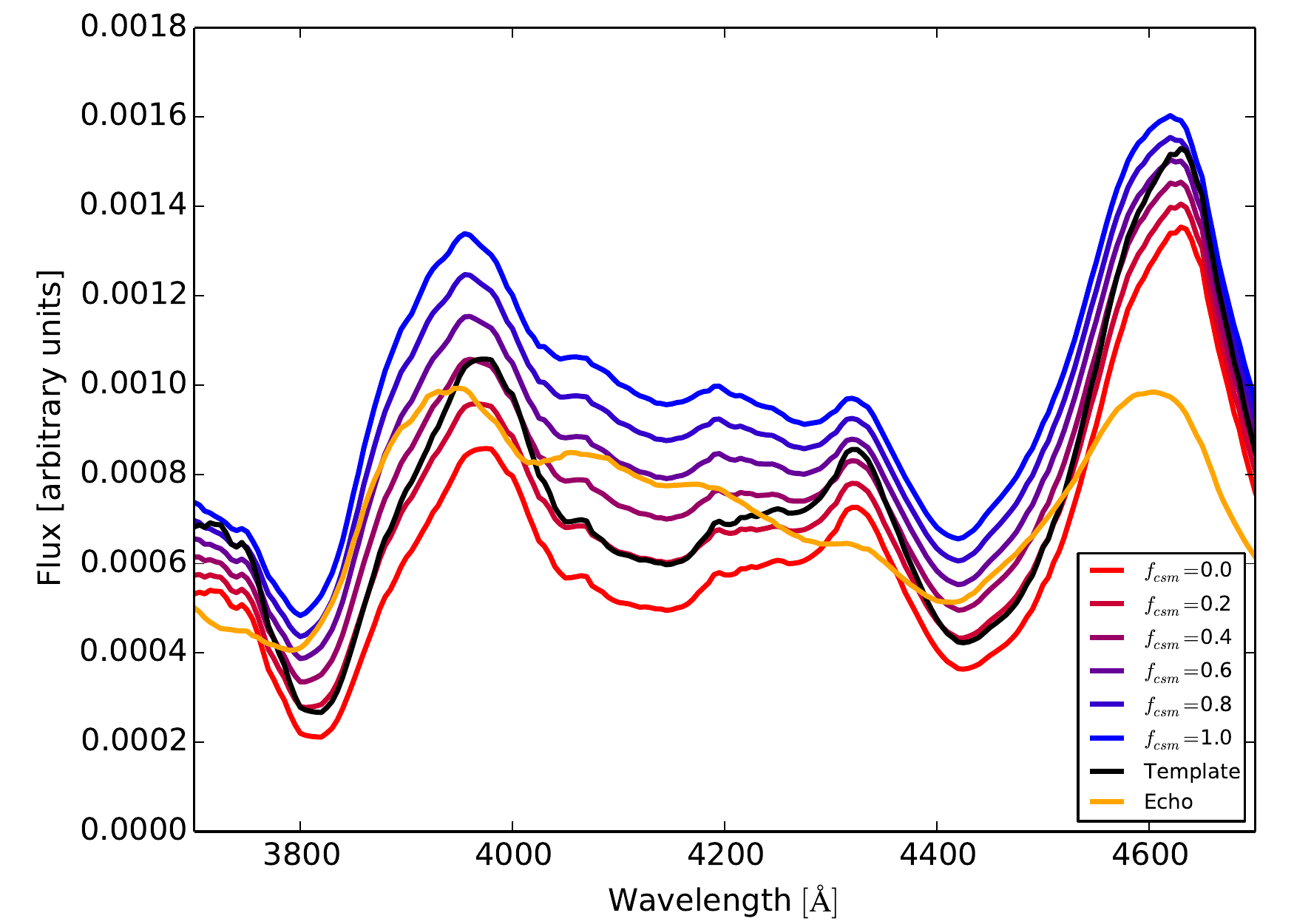}
\caption{Simulated spectra at 50 days past maximum light with total
  extinction $A_{V}=0.5$ and $R_{V}=3.1$, and LEs due to CSM. The
  different lines represent different fractional amounts of CSM
  ($f_{\mathrm{CSM}}$) with a radius of 0.05 pc and a fixed total
  column density. The black line represents the original spectrum at
  50 days past maximum light without extinction or LEs, while in orange
  the LE spectrum is shown in an arbitrary scale. All models are
  normalized to $V$-band flux.}\label{echo_zoom}
\end{figure}

In this simulation the main signature due to LEs is found near 4100
\AA. This can be seen in Figure \ref{echo_zoom}, where the shape of
the spectrum gets considerably modified in the LE scenario (purple and
blue lines) producing a characteristic signature. On the other hand,
in the pure extinction scenario if the column density is reduced
(black line), it produces just a smooth change in the spectrum
compared to the same spectrum with extinction (reddest line). In
particular, the shape of the feature near 4100 \AA \ will not be
affected.


These features change with the distance $R$ between
the CSM and the SN. Reducing the distance is analogous to making
$tr=t$ (see eq. \ref{tr}) obtaining the result of a pure extinction
model where
$S(t,\lambda)=\omega(\lambda)(1-X^{f_\mathrm{CSM}})f^0(t,\lambda)$. On
the other hand, increasing $R$ makes the light echoes more diluted in
time reducing the ratio between the LE and the intrinsic flux
$f$. Hence, there is an optimum $R$ near 0.05 pc in which LEs can
affect the colors and spectra during the Lira phase.

\begin{figure}[h!]
\centering
\includegraphics[trim=0.0cm 0.0cm 0.0cm 0.0cm, clip=true,width=1.0\linewidth]{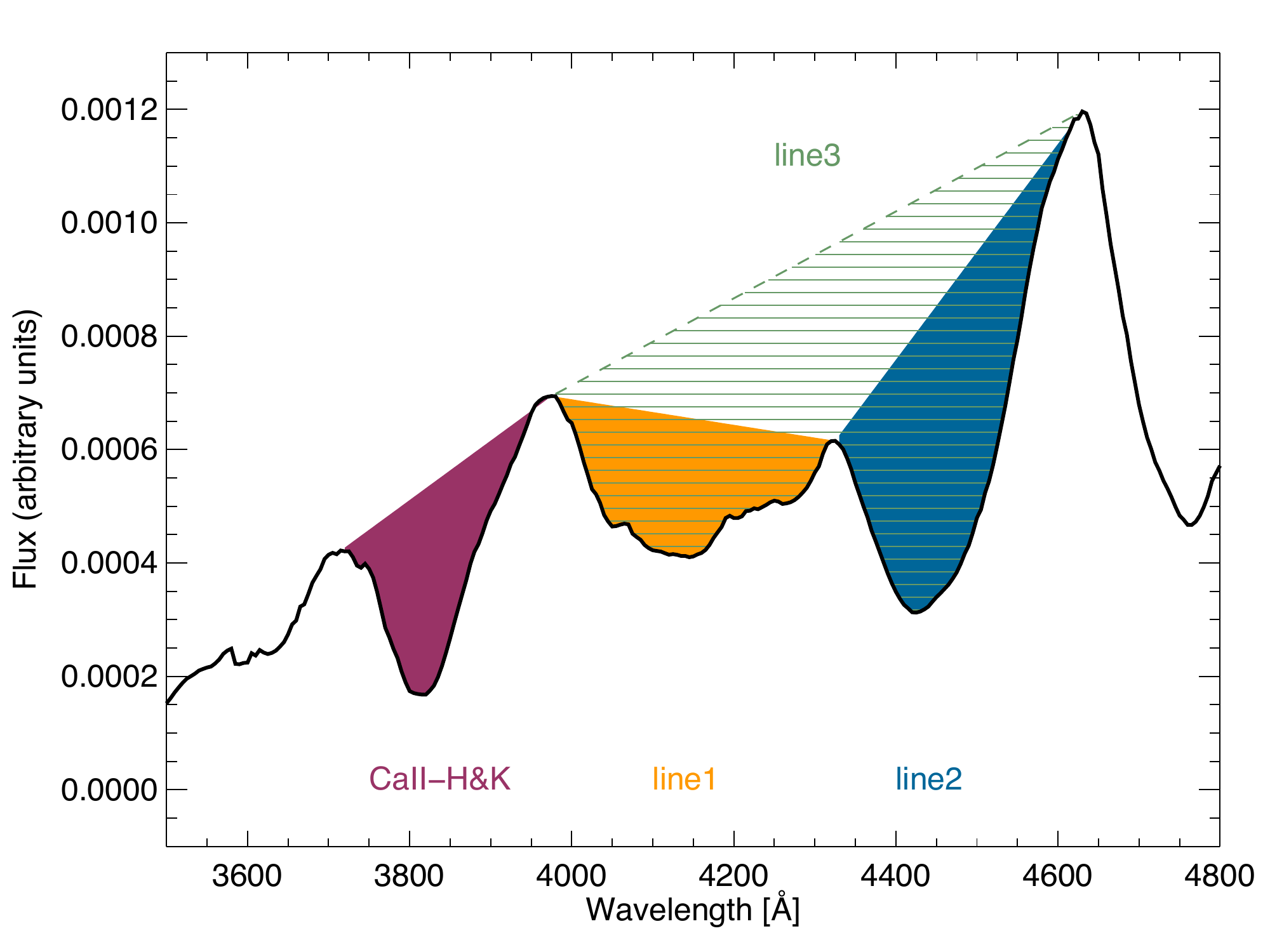}
\caption{Features considered in the analysis as possible indicators of
  LE signatures: \ion{Ca}{2} H\&K complex at 3500-4100 \AA, ``line 1''
  at 3800-4400 \AA, ``line 2'' at 4250-4800 \AA \ and ``line 3'' at
  3800-4800 \AA, the union of line 1 and line 2. Lines 2 and 3 come
  from a blend of \ion{Mg}{2} and \ion{Fe}{2}. The shaded regions
  indicate the pW for \ion{Ca}{2}H\&K, lines 1 and 2 and the dotted
  line shows the pseudo-continuum of line 3.}\label{lines}
\end{figure}

To further investigate the LE effects on spectra, we focus on the
range between 3000-5000 \AA, particularly on the absorption lines and
their related measurable quantities such as the equivalent width and
the slope of the continuum. We define five characteristic features,
four of them that are presented in Figure~\ref{lines}: the \ion{Ca}{2}
H\&K complex between 3500-4100 \AA, and two absorption features
originating mainly from a blend of \ion{Mg}{2} and \ion{Fe}{2} at
3800-4400 \AA\, (``line 1'') and 4250-4800 \AA\, (``line 2''); these two
together form another larger feature at 3800-4800 \AA\, (``line 3''),
equivalent to the feature pW3 in \citet{Folatelli13}. Additionally, we
use the line feature around the Si II absorption around 5800-6300\AA\,
(``line 4")

\begin{figure}[h!]
\centering
\includegraphics[trim=0.0cm 0.0cm 0.0cm 0.0cm, clip=true,width=1.0\linewidth]{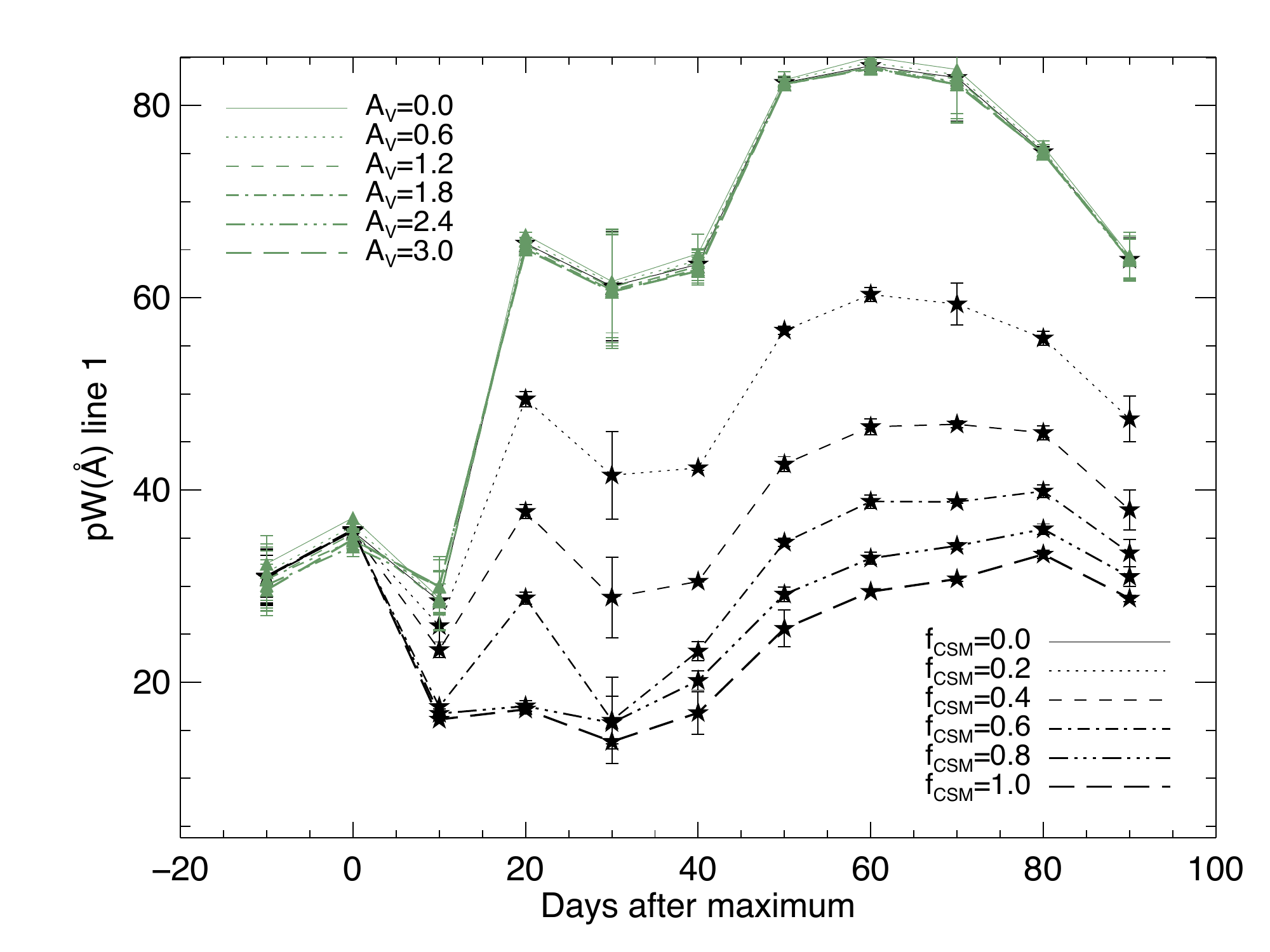}
\caption{Pseudo-equivalent width evolution of the line 1 feature
  for simulated spectra. The green lines represent the evolution of
  line 1 with different amounts of extinction ($A_V$), while in
  black with different CSM amounts ($f_{\mathrm{CSM}}$), with fixed
  radius of 0.05 pc and a total column density or $A_V=1.0$.}\label{ews}
\end{figure}

For all LE and extinction models we measure pseudo-equivalent widths
(pW), i.e. with a pseudo-continuum in a similar fashion to
\citet{Garavini07,Bronder08,Folatelli13}. We use a semi-automatic
algorithm that searches for the pseudo-continuum at specific regions
defined in Table~\ref{features}. The algorithm also calculates the
slope of the pseudo-continuum, which we find to be another good
indicator of LEs. In figure~\ref{ews} we show the predicted evolution
of the pWs of line 1 for a SN with pure extinction (green lines) and
with LEs due to CSM (black lines). The pWs vary significantly during
the Lira phase for LE models with different CSM fractions,
$f_{\mathrm{CSM}}$, whereas they do not for different amounts of
extinction. This effect is strongest at larger CSM distances of
0.05-0.25pc. The difference between the scenario with extinction and
with LEs is still clearer and less unbiased by the SN intrinsic pWs if
we normalize by the pW at maximum light (hereafter pW ratio).

From the five pW ratios and respective slopes, we find that the best
candidate to be a CSM indicator is line 1, the other lines show less
differences in their evolution between different extinctions and CSM
scenarios.

\begin{table}[h!t]
 \centering
   \caption{Feature definitions for pseudo-equivalent width and pseudo-continuum.} 
 \begin{tabular}{c|c|c}
    \hline
    \hline
    Feature & Blueward limit range (\AA) & Redward limit range (\AA) \\ 
    \hline
    \ion{Ca}{2} & 3500 - 3800 & 3900 - 4100 \\
    line 1 & 3800 - 4100 & 4250 - 4400 \\
    line 2 & 4250 - 4400 & 4400 - 4800 \\
    line 3 & 3800 - 4100 & 4400 - 4800 \\
    line 4 & 5800 - 6000 & 6100 - 6300 \\
    \hline
  \end{tabular}
\label{features}
\end{table}

\section{Comparison with observed spectra}\label{analysis}


LEs are faint, and therefore they do not contribute significantly to
the SN spectra around maximum. Following this reasoning, we first
analyze the reddening at maximum light, to determine the total
extinction or $A_{V}$ due to CSM and/or ISM using standard extinction
laws (see \S \ref{extinctionlaw}). Then, analyzing the best fits to
the spectra at later epochs we can discern if it is compatible with
the extinction found at maximum light or if it is necessary to include
LEs or, alternatively, reduce the amount of dust in the line of sight.

\subsection{Data}
\label{data}
The spectra we use were taken from the Carnegie Supernova Program CSP
\citep{Folatelli13}, public data of the Center for Astrophysics CfA
\citep{Blondin12}, the Berkeley Supernova Ia program BSNIP
\citep{Silverman13} and The Online Supernova Spectrum Archive SUSPECT
(see table \ref{t:suspect}). We analyzed only the subset of SNe that
were already classified as fast or slow Lira decliners during the Lira
Law phase in F13.

\begin{center}
\begin{table*}[ht]
\caption{Nearby SNe Ia spectra used in this analysis (besides the data from CSP, CfA and BSNIP) from The Online Supernova Spectrum Archive (SUSPECT).}
\centering
\label{t:suspect}
\hfill{}
\begin{tabular}{lc}
\hline
\hline
\noalign{\smallskip}
Name & Sources \\
\noalign{\smallskip}
 \hline
SN 2005hk	&	\citealt{2007PASP..119..360P};	\citealt{2012AJ....143..126B}; \citealt{2006PASP..118..722C}	\\
SN 1999ac	& \citealt{2005AJ....130.2278G} \\
SN 1998aq	&	\citealt{2003AJ....126.1489B} \\
SN 2005cf		&	\citealt{2007AA...471..527G};		\citealt{2009ApJ...697..380W}; \citealt{2009ApJ...700.1456B}; 	\citealt{2007AIPC..937..311L} \\
SN 2003du &	\citealt{2007AA...469..645S}	\citealt{2005AA...429..667A}	\citealt{2005ASPC..342..250G} \\
SN 2005am &	\citealt{2007AIPC..937..311L}\\
SN 2006X	& \citealt{2009PASJ...61..713Y}; 	\citealt{2008ApJ...675..626W}	\citealt{2011Sci...333..856S}	\\
SN 1999aa &	\citealt{2004AJ....128..387G}\\
SN 2002bo &	\citealt{2004MNRAS.348..261B}	\\
SN 2000cx  &	 \citealt{2001PASP..113.1178L} \\
SN 1994D	& \citealt{1996MNRAS.278..111P};	\citealt{1998AJ....115.1096G}	\\
SN 2003cg &	\citealt{2006MNRAS.369.1880E}	 \\
\noalign{\smallskip} \hline  \noalign{\smallskip}
\end{tabular}
\hfill{}
\end{table*}
\end{center}

First, the spectra are corrected for Milky Way extinction using the
values from \citet{Schlafly2011} and deredshifted to rest-frame . Then
we smooth the spectra using a non-parametric fit with a velocity
window of 1000 km/s and a wavelength regridding of 5 \AA. We also
compute the dispersion of each original spectrum with respect to its
smoothed version to estimate the noise in our smoothed spectra. We
normalize each smoothed spectrum to the same $V$ band flux by
numerically convolving the spectrum with the filter transmission
function \citep{Bessell90}, in order to put all spectra in the same
scale and be able to compare the shape and features of the spectra
instead of the absolute fluxes, which are difficult to calibrate
precisely.  We also adjust the shape of the spectra to match the
observed colors interpolated to the given epoch \citep{Hsiao07}, in
order to have spectra consistent with the available photometry and
with the previous work in F13.



We analyze the different SN spectra at maximum light and during the
Lira law phase at 5 different epochs or time windows centered at 40,
50, 60, 70 and 80 days after maximum light with a width of 10 days. To
have a single representative spectrum at every time window per SN, we
make weighted average spectra with the available spectra of each
SNe. For more details about this weighted averages check \S
\ref{A:average_spectra}.

\subsection{Template spectra}
\label{templates}

F13 showed that slow Lira decliners present weaker equivalents width
(EW) of blended Na I D1 \& D2 narrow absorption lines, while fast Lira
decliners have stronger EWs and redder colors independent of
environmental factors. One possible interpretation of these results is
that fast Lira decliners have CSM that produces LEs or that they have
nearby dust that is sublimated in time. In order to test these
hypothesis and analyze the spectra of fast Lira decliners with our
models, we compare them with some standard unreddened and non-evolving
CSM spectral time series given in this case by the slow Lira
decliners. For this, we construct different template spectra that
cover the intrinsic SN Ia variability at different
epochs. \citet{Chotard11} showed that most of the intrinsic
variability of spectra in SNe Ia at 2.5 days after maximum light can
be characterized with the equivalent width (EW) of Si II 4131 \AA\ and
Ca II H\&K. We follow the approach and use these two lines and a
stretch parametrization to describe the shape of the light curve and
defined as a factor multiplying the time axis \citep{Perlmutter97,
  Goldhaber01} to construct different templates. However, we can not
measure the EWs of the Ca and Si lines in all of our SNe Ia. Given
than the EW of Si II 4131 is correlated with the light curve stretch
parameter, we decide to use epoch and stretch as our main variables to
construct templates accounting for the intrinsic variability in SN Ia
spectra. \\

\begin{figure}[h!]
\centering
\includegraphics[width=1.0\linewidth]{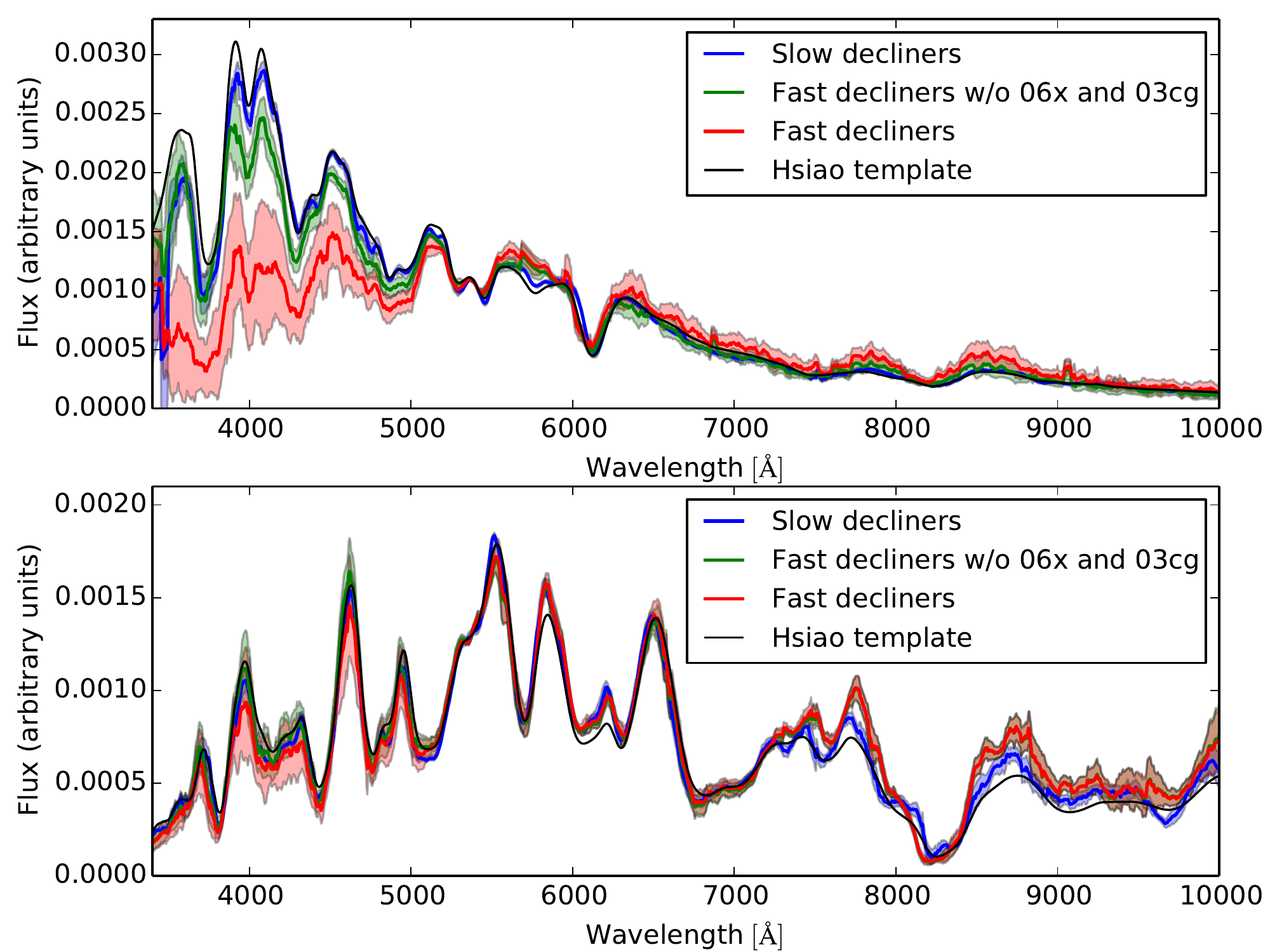}
\caption{Bootstrap templates for slow and fast Lira decliners SNe at
  maximum light (upper panel) and 50.0 days after maximum light(lower
  panel), for a stretch of 0.98. In blue is the $B-V$ slow Lira
  decliner template , in red (green) is the $B-V$ fast Lira decliner
  template including (without) SN 2003cg and SN 2006X. In black line a
  template by \citet{Hsiao07} is presented. The filled contours
  represent the one standard deviation regions based on the dispersion
  in the spectra.} \label{bootstempslowfast}
\end{figure}

To ensure that these average templates are not heavily biased by few
extreme SNe, we perform a bootstrap simulation, i.e., we constructed
100 different templates using random sets from the original. For more
details see \S \ref{A:template}. In figure \ref{bootstempslowfast} we
compare the bootstrap templates for fast and slow Lira decliner
spectra at maximum light and 50 days later, and one standard deviation
($\sigma$) regions. At maximum light there is a clear difference
between fast and slow Lira decliners near 4000 \AA, the average
difference is about 1.4$\sigma$ between 3400 - 10000 \AA \ and
0.8$\sigma$ if we do not consider SN~2006X and SN 2003cg. Between 3400
- 5500 \AA \ the difference is about 2.3$\sigma$ and 1.4$\sigma$ if we
do not consider SN 2006X and SN 2003 cg. But this difference after 50
days decreases to 0.8$\sigma$ between 3400 - 10000 \AA \ (0.9$\sigma$
without SN 2006X and SN 2003cg) and 0.5$\sigma$ between 3400 - 5500
\AA \ (0.6$\sigma$ without SN2006X and SN2003cg). We also find that
the dispersion among fast Lira decliners is larger than in our slow
Lira decliners sample. These results confirm what was found in F13,
and we stress the fact that they are valid irrespective of warping the
spectra to the observed colors. Finally, we point out that the
template of the slow Lira decliners is quite similar to the template
by \citet{Hsiao07} suggesting that this is the group of more
``normal'' unreddened SNe~Ia.

\subsection{Extinction law at maximum light} 
\label{extinctionlaw}
To find an extinction law for each fast Lira decliner at maximum
light, we use as unreddened reference a slow Lira decliner template
representing the same epoch and stretch, and the extinction law
described in \citet{Fitzpatrick90}, adopting the mean values for the
parameters found in \citet{Fitzpatrick99} and leaving the visual
extinction $A_{V}$ as free parameter.

We fix the total-to-selective extinction ratio $R_{V}$ at the
standard Milky way value of 3.1. A discussion on this is
presented in \S\ref{discussion}. For each fast Lira decliner average
spectrum we fit $A_{V}$ and a normalization parameter $C_{1}$, that
corrects the fact that our spectra are normalized to their V-band flux,
minimizing a chi-square function (for more details see \ref{A:ext}).

To analyze global differences between both subsets of SNe we fit the
extinction law of \citet{Fitzpatrick99} to the bootstrap templates of
fast Lira decliners using as reference the bootstrap templates of slow
Lira decliners presented in Figure \ref{bootstemp}. We find at
maximum light an extinction with $A_V = 0.64 \pm 0.04$ and $R_V = 3.0
\pm 0.2$ when we include the highly reddened SN 2006X and SN
2003cg. When we exclude them we obtain $A_V = 0.24 \pm 0.01$ and $R_V
= 3.3 \pm 0.1$. However, during the Lira law we obtain an extinction
with $A_V = 0.17 \pm 0.01$ and $R_V = 3.3 \pm 0.1$ significantly lower
than at maximum light and when we exclude these SNe we obtain $A_V =
0.04 \pm 0.01$, $R_V = 3.6 \pm 0.1$. These results recover some of the
results in F13, fast Lira decliners show a greater extinction at
maximum light, which starts to decrease during the Lira law. Moreover,
their $R_V$ slightly increases over time.

\subsection{Light Echo fit}
Using the model described in \S\ref{model} plus the values for the
dust albedo $w$ and the degree of forward scattering $g$,
corresponding to dust with $R_{V}=3.1$, we fit the LE model to
observed spectra of fast Lira decliners in the Lira phase. The parameters to fit are
$f_{\mathrm{CSM}}$ and a normalization constant $C_{2}$ to correct for
the fact that our spectra were normalized by their flux in the V
band.

The function to minimize for each average spectrum $i$ is
\begin{equation}
\chi^{2}_{i}=\sum_{\lambda}\frac{(f^{i}(t,\lambda)-C_{2}F(t,f_{\mathrm{CSM}},\lambda))^{2}}{\delta f^{i}(t,\lambda)^{2}+ (C_{2}\delta F(t,f_{\mathrm{CSM}},\lambda))^{2}} \label{eq:chi2echo}
\end{equation}
where
\begin{small}
\begin{equation}
F(t,\lambda)=f^{0}(t,\lambda)10^{-0.4A(\lambda)}+S(t,\lambda)10^{-0.4A(\lambda)(1-f_{\mathrm{CSM}})} 
\end{equation}
\end{small}
To evaluate $S(t,\lambda)$ we use eq. \ref{eq: echo} in a slightly
different version because we do not have the intrinsic fluxes in our
spectra as they were previously normalized (for more details, see
\ref{A:LEfit}).

\section{Results} \label{results}
\subsection{Extinctions}\label{extresults}
Out of 31 individual spectra from different fast Lira decliners at
maximum light, minimizing the $\chi^2$ of eq. \ref{eq:chi_ext} we
obtain 24 SNe with positive $A_{V}$ values and 7 with nonphysical
negatives values. Those SNe are excluded from our sample for the LE
analysis as they present less extinction than the template at maximum
light, making the posterior fit of our LE model impossible. This may
be caused by some SNe in the sample of slow Lira decliners having non
negligible host extinction. At maximum light the mean $A_{V}$ is 0.44
while excluding the highly reddened SNe, SN~2006X and SN~2003cg, it
decreases to 0.27. But during the Lira law the amount of extinction
starts to decrease to values near zero. In Table \ref{deltaAv} the
mean differences in $A_V$ per SN are presented with respect to the
value found at maximum light excluding those SNe with negative $A_{V}$
at that epoch. This result confirms that fast Lira decliners are more
extincted at maximum light than slow Lira decliners, and during the
Lira law both groups become more similar, as shown in F13. This is
also valid without the warping of the spectra to the observed
colors. However, there are a few SNe that show a positive $\Delta
A_{V}$. This could be produced by an artifact in our templates or a
bad extinction law fit.

\begin{figure}[h!]
\centering
\includegraphics[width=1.0\linewidth]{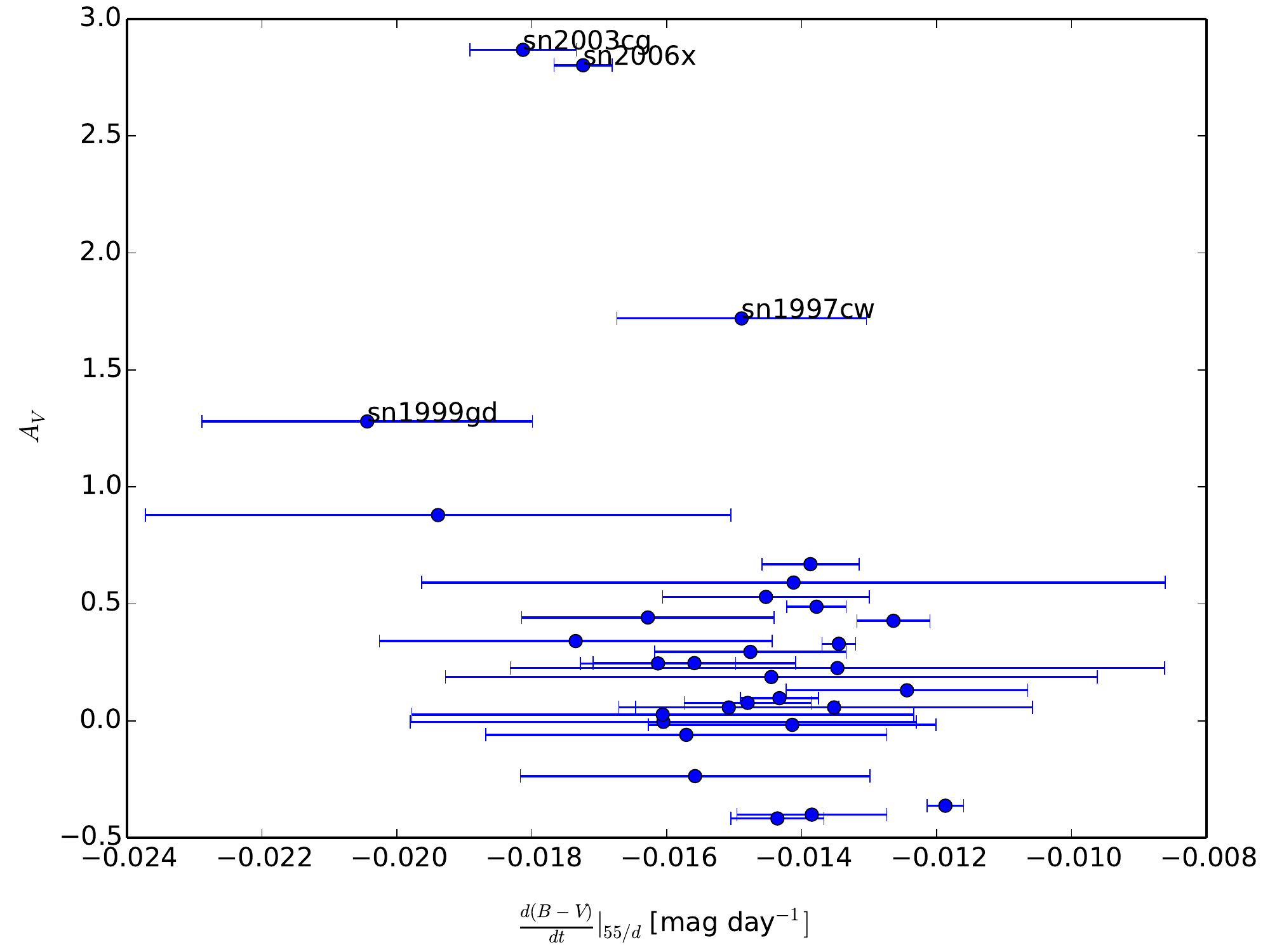}
\caption{$A_{V}$ fit results at maximum light vs $B-V$ slopes of the Lira law from F13.\label{Avs}}
\end{figure}

We find a small correlation between the $A_{V}$ values found at
maximum light and the $B-V$ slope of the Lira law, with a Pearson's
correlation coefficient of -0.5 (see figure \ref{Avs}), which is
consistent with the analysis done in F13.

\begin{center}
\begin{table*}[ht]
\caption{Mean and uncertainty of $\Delta A_{V}=A_{V}(t)-A_{V}(max)$ for fast Lira decliner SNe at different epochs.}
\centering
\label{deltaAv}
\begin{tabular}{c|c|c|c|c}
\hline
\hline
 40 days & 50 days & 60 days & 70 days & 80 days \\
\hline
$-0.26\pm 0.07$ & $-0.19\pm 0.09$ &$-0.54 \pm 0.11$ & $-0.21 \pm 0.17$ & $-0.29 \pm 0.18$ \\
\hline 
\end{tabular}
\end{table*}
\end{center}

\begin{center}
\begin{table*}[ht]
\caption{Comparison of ISM, LE and DS models with overall spectra: fraction of fast Lira decliner SNe with positive $\mathrm{A}_{\mathrm{V}}$ at maximum light that have a BIC value which favors the LE or DS models vs the other models at different time windows.}
\centering
\label{t:echofit}
\begin{tabular}{lccccc}
\hline
\hline
\noalign{\smallskip}
& 40 days & 50 days & 60 days & 70 days & 80 days \\
\noalign{\smallskip}
\hline
Number of SNe  &16 & 11 & 8 & 10 & 7\\
LE vs ISM                & 0.50 & 0.36 & 0.63 & 0.5 & 0.57 \\
LE vs ISM \& DS     & 0.19 & 0.18 & 0.13 & 0.2 & 0.14\\
DS vs ISM \& Echo & 0.44 & 0.27 & 0.63 & 0.3 & 0.43 \\
\noalign{\smallskip} \hline  \noalign{\smallskip}
\end{tabular}
\end{table*}
\end{center}

\begin{center}
\begin{table*}[ht]
\caption{Comparison of ISM, LE and DS models with specific spectral regions: fraction of fast Lira decliner SNe with positive $\mathrm{A}_{\mathrm{V}}$ at maximum light that have a BIC value which favors the LE or DS models vs the other models at 50 days after maximum light for different wavelength regions.}
\centering
\label{t:echofitranges}
\hfill{}
\begin{tabular}{lccccc}
\hline
\hline
\noalign{\smallskip}
& 3900-10000 \AA & 3000-6500 \AA & 6500-10000 \AA & 3600-4800 \AA & 5900-6400 \AA \\
\noalign{\smallskip}
 \hline
LE vs ISM               & 0.36  & 0.45 & 0.45 & 0.27 & 0.18 \\
LE vs ISM \& DS     & 0.18  & 0.36 & 0.45 & 0.27 & 0.18  \\
DS vs LE \& ISM & 0.27 & 0.18 & 0.0 & 0.0 & 0.0 \\
\noalign{\smallskip} \hline  \noalign{\smallskip}
\end{tabular}
\hfill{}
\end{table*}
\end{center}

\subsection{Light echoes} \label{Res: LE}
Once an $A_{V}$ per fast Lira decliner SN is found, we fit our
light echo model to the different fast Lira decliner spectra to
search for spectral evidence of LEs. We look for the minimum of
eq. (\ref{eq:chi2echo}) by varying $f_{\mathrm{CSM}}$ and $C_{2}$. We
fix $R_{V}$ to 3.1 and the radius of the CSM shell to 0.05 pc (more
on this in \S\ref{discussion}).

 We also fit the data with two other models to compare the goodness of
 fit of the light echo model applied to the observations. The three
 models that we compare are:
\begin{enumerate}
 \item The light echo model in which we fit for $f_{\mathrm{CSM}}$ and
   $C_{2}$. It also includes ISM extinction to get a
   total extinction consistent with the amount of extinction found at
   maximum light.
 \item A pure ISM model that just uses a late time spectral template with
   the same extinction found at maximum light. In this scenario we
   just fit a normalization constant $C_{1}$.
 \item A dust sublimation (DS) model consisting of a new pure
   extinction fit during the Lira law, with a lower $A_V$ value than
   the one found at maximum light.
\end{enumerate}

Then we calculate the Bayesian Information Criterion (BIC) parameter
in order to compare the goodness of the three models applied to the
same data. The BIC parameter is obtained as
\begin{equation}
\mathrm{BIC}=\chi^{2}+k\cdot \mathrm{ln}(n)
\end{equation}
where $k$ is the number of free parameters to fit in each model and
$n$ is the number of data points or wavelengths with measured flux
values. The best model is the one with the lowest BIC. It is a
combination of $\chi^2$ plus a function that penalizes having too
many free parameters overfitting the data. With the BIC parameter we
judge whether the LE model is able to explain the observed spectra
better than a simple extinction or DS model. The results of the fits
to the entire spectra and comparison between the three models are
shown in Table \ref{t:echofit}.

According to our results, the LE model works better than using just
the extinction law found at maximum light in ~50\% of the cases. But
when we compare it to the dust sublimation model, the fraction of
favorable cases drops to values lower than 20\%, giving larger weight
to a scenario where $A_V$ simply decreases with time.

We also fit the three models to the spectra in particular wavelength
ranges: blue region (3000 - 5000 \AA); red region (6500 - 10000 \AA);
signature I (3600 - 4800 \AA) and signature II (5900 - 6400 \AA). The
results are shown in Table \ref{t:echofitranges}. The analysis in
wavelength ranges is not favourable for the DS model, and the simple
extinction found at maximum light works better than the LE model in
most cases. However, this can be explained by the fact that when we
look at particular regions of a spectrum a degeneracy appears between
the normalization and the extinction, as there is not enough
wavelength range to anchor the reddening, obtaining a better BIC with
the constant extinction model than with the DS model which has more
parameters.


For each late-time SN spectrum, we derive the value for the CSM
fraction $f_{\mathrm{CSM}}$. Multiplying this value with the $A_{V}$
found at maximum light we can infer the fraction of the extinction due
to CSM under the hypothesis of LE. For the four SNe with $A_V>1$ at
maximum light, we obtain that their $A_V$'s due to CSM are lower
than 0.5, validating our single scattering assumption.

\begin{figure}[h!]
\centering
\includegraphics[width=1.0\linewidth]{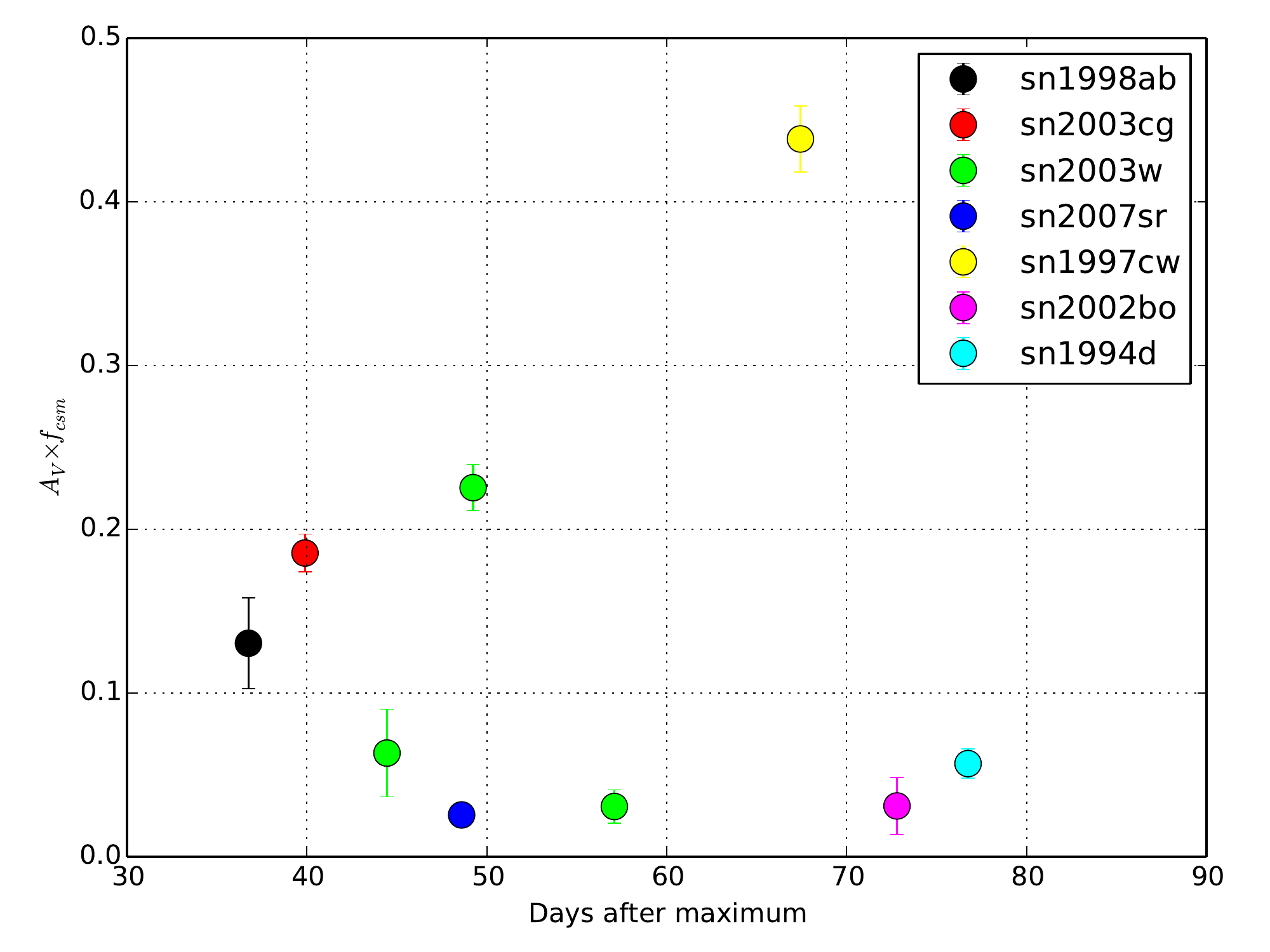}
\caption{$A_{V}\times f_{\mathrm{CSM}}$ vs time since maximum
  light.\label{fcsm} Only cases in which the LE model works better
  than the pure extinction and DS models are shown. The error bars
  represent three standard deviation errors on the parameters that
  minimize the $\chi ^2$ function under our LE model assumption. }
\end{figure}

In Figure \ref{fcsm} we show $f_{\mathrm{CSM}}\times A_{V}$ for
different SNe at different epochs. Only epochs in which the LE model
was the best according to its BIC value are plotted. If the CSM were
not disturbed by the supernova we would expect $
f_{\mathrm{CSM}}\times A_{V}$ to stay almost constant for each
SN. Unfortunately, only one object, SN 2003W, has more than one
late-time spectrum consistent with LE to allow us to perform this
test. The three resulting $f_{\mathrm{CSM}}\times A_{V}$ values range
between 0.03-0.23, which we consider a satisfactory agreement
considering the simplifications of the LE model.


The number of SNe for which the LE model has a favorable BIC in
comparison with the ISM and DS models is small (7 SNe), and just
SN2003W appears in more than one epoch. Even in such cases, a visual
inspection of the spectra does not reveal the signatures expected from
LEs. This poses serious questions on the LE scenario.

\subsection{Line comparison}

\begin{figure}[h!]
\centering
\includegraphics[trim=0.0cm 0.0cm 0.0cm 0.0cm, clip=true,width=1.0\linewidth]{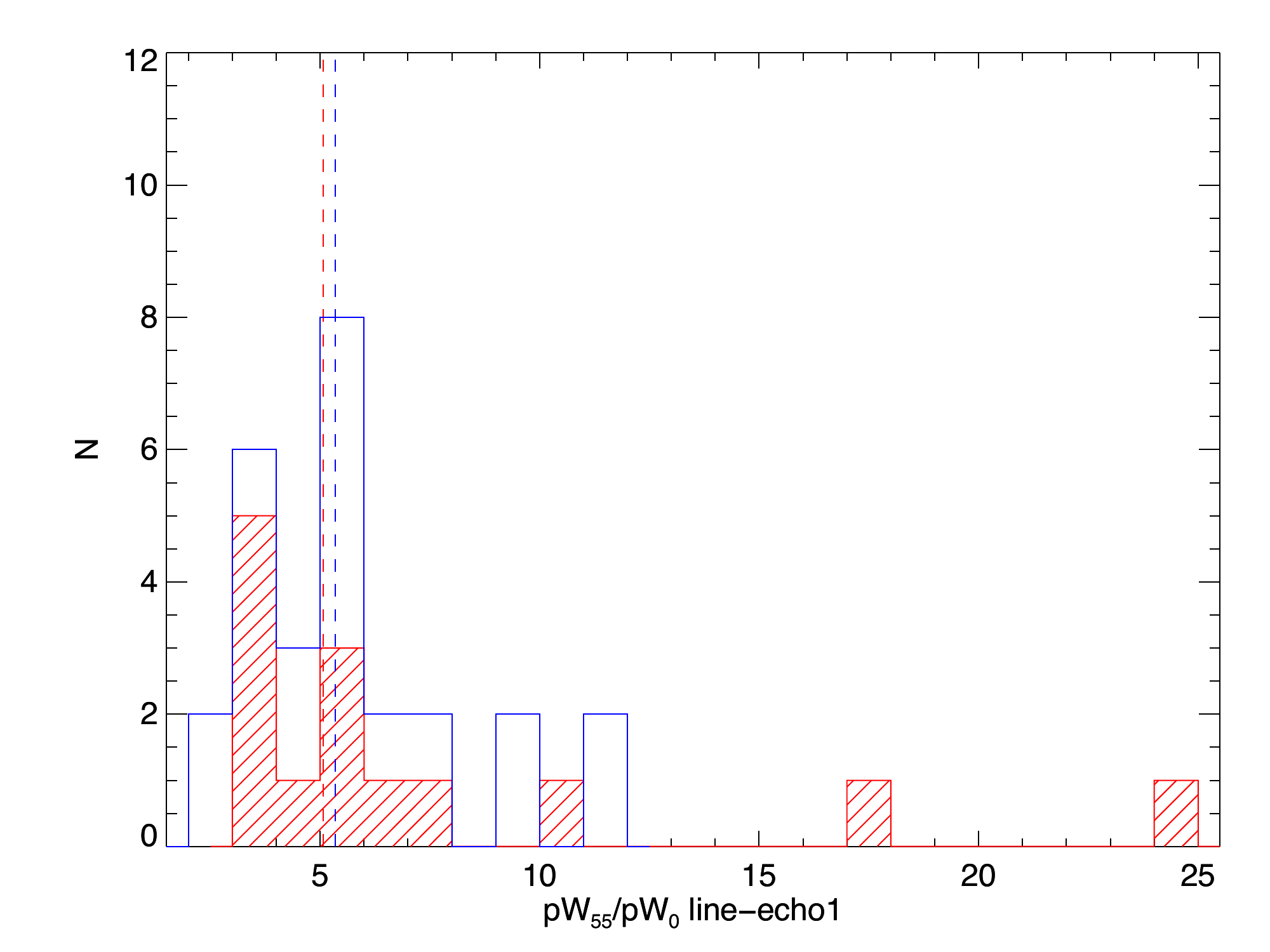}
\caption{Histograms of the ratio of pseudo-equivalent width at maximum
  light and at 55 days past maximum light for the diagnostic echo
  line 1 for the sample of slow (blue) and fast (red) Lira
  decliners. Vertical dashed lines show the median of the population:
  5.4$\pm$1.70 for slow and 5.7$\pm$1.57 for fast Lira decliners,
  respectively. The KS test for these distributions gives a probability
  of 99\% of being drawn from the same distributions}\label{EWratio}
\end{figure}

Another way to test our LE model is to use the line diagnostics
presented in section~\ref{predictions}. For this we calculate for all
our spectra the pW and pseudo-continuum slope for the four diagnostic
lines previously defined, in the same way we measure them in the
simulated spectra. We obtain these features at all available epochs
and also normalize the pW curves by their values at maximum light in
order to study their evolution. Finally we compare the results for the
slow and fast Lira decliners with the simulated spectra. Apart from
some slight differences between the two samples, we find that in
general the populations are consistent within the errors. Both groups
of SN Ia show similar trend and dispersion in their evolution. As an
example, Figure \ref{EWratio} shows the distribution of pW at 55 days
after maximum light for the two SN samples. In the context of the LE
model, we would expect both a difference in the dispersion of pW´s and
their mean value, neither of which is observed.


To measure possible statistical differences we perform a K-S
 test. We find that for line 1 the two populations, slow and
fast Lira decliners, have a 99\% probability of being drawn from
the same distribution. We check all of the other lines for which we
obtain low K-S values, yet we do not find any hint of LE
signatures among fast Lira decliners according to our predictions (KS
values between 0.2 and 0.99) . The pW distributions are consistent for
both populations during the Lira phase. However, some pseudo-continuum
slopes show very low KS values (lower than 0.01). This
difference in both SN samples can be explained by the greater color
dispersion of fast Lira decliners which also tend to be redder at
maximum light, as shown in F13. As a matter of fact, the trend goes
in the opposite direction of what is predicted by our LE models.

Although we do not find a correlation between the evolution of lines
and fast Lira decliners, our CSM model predictions, especially for
line 1, can be important to diagnose the presence of LEs in other
samples of SNe~Ia.



 
\section{DISCUSSION} \label{discussion}

\subsection{Extinction laws and $R_V$} \label{dis:ext}



To calculate the extinction at maximum, we also use other reddening
laws such as \citet{Cardelli89} with the inclusion of
\citet{ODonnell94}, obtaining similar results to \citet{Fitzpatrick99}
when keeping $R_{V}$ fixed. We also explore the reddening law proposed
in \citet{Goobar08}, but we discard it for two reasons: the reddening
law becomes degenerate with our normalization constant $C_1$ and also
because it accounts for the observed reddening in the context of CSM
without considering the evolution of the radiated spectrum. In
contrast, we aim to find an intrinsic reddening law and the time
evolution of the spectrum caused by LEs.


To be consistent in our predictions and in our fits we used extinction
laws with a fixed $R_V$ of 3.1 as we use standard MW values for the
albedo and phase function of interstellar dust. Therefore our analysis
is restricted to a statistical point of view more than the study of
particular cases. The $R_{V}$ value can vary depending on the
properties of the dust (e.g. grain size distribution and composition)
and seems to be different from the MW in the line of sight towards
some SNe Ia \citep{Mandel11, Burns14}. It is not very clear what range
of values are consistent with circumstellar dust surrounding a SN. It
is very important to differentiate in extinction laws analysis between
the intrinsic $R_V$, which comes from the dust properties, and the
observed $R_V$ when a pure extinction law is assumed, omitting more
complex interactions. A future improvement in our model is to compute
and use the specific opacities, albedo and phase function given any
dust grain size distribution and composition.

\subsection{Light echo models}

We have explored the possibility of detecting LEs due to CSM in SN Ia
spectra. Our results show that LEs are not a global phenomenon on fast
Lira decliner SNe during the Lira law phase. Even though we find that
for $\sim$50\% of the spectra the LE model works better than the
extinction law derived at maximum light, the number of favorable cases
drop to values near 15\% when we compare them with the DS models.
                               

We also fit the LE model using a CSM radius $R$ of 0.01 and 0.25
pc, instead of 0.05pc. With the smallest $R$ we obtain a lower fraction of favorable
cases for the LE model compared to the original results with $R=0.05$
pc, even lower than 50\% when the LE model is compared with just the
pure extinction model (ISM). On the other hand, when we fix $R=0.25$
we recover almost the same results than the original CSM
scenario. Therefore, if CSM is present, larger radii of 0.05-0.25pc are favoured. In principle, it is possible to fit at the same time $R$,
$f_{\mathrm{CSM}}$ and $C_2$, but this is computationally expensive
and could overfit the data.

Our CSM model consists of an isotropic spherical shell in the limit of
negligible thickness. We did not consider multiple scattering, which
at optically thin scenarios is negligible.  Nevertheless, we know that
multiple scattering could become important at optical wavelengths when
the optical depth is larger than 1, i.e. $A_{V} \gtrsim 1$. Therefore
we are unable to analyze SNe with an expected $A_{V}$ due to CSM
larger than one, but according to our results none of the SNe in our
sample presented an $A_{V}$ due to CSM larger than 1 (including
SN2003cg and SN2006x). A model that includes multiple scattering is
necessary to predict the effect on the light curves and spectra when
an optically thick CSM is present \citep{Amanullah11, Patat05}.

The CSM geometry may probably be different from a spherical shell,
e.g. non-isotropic disk or ring geometries formed from a planetary
nebula have been proposed recently to model time variable Na I D
absorption \citep{Soker2014}. The predicted LE signatures might vary
depending on the CSM geometry and orientation to the observer.



For an optically thin shell, a rough estimate of the total dust mass
in our models can be obtained for a given CSM radius, an $A_V$ and a
typical ISM dust opacity:
\begin{small}
\begin{equation}
M_d=6.4\times10^{-5} \left(\frac{R}{0.01 \ \mathrm{pc}} \right)^2 \left( \frac{A_V}{0.1} \right) \left( \frac{\kappa_V}{8.55\times10^3 \mathrm{cm}^2\mathrm{g}^{-1}}\right)^{-1}  \mathrm{M}_{\odot} \
\end{equation}
\end{small}

This is a very high mass in form of dust for a CSM. However, we are
assuming a spherical shell and a specific opacity corresponding to
interstellar extinction in the MW. If we consider a different geometry
for the CSM or a larger specific opacity, i.e. smaller grains, the
inferred mass will vary by orders of magnitude.

 

\subsection{Light echo effects on the light curves} \label{LE_lightcurves}

Using Monte Carlo simulations, \citet{Amanullah11} found that
different radius or $E(B-V)$ for the CSM could affect the $B-V$
evolution during the Lira law phase in different ways. In order to
test the hypothesis made in F13 that LEs could increase the $B-V$
decline rate during the Lira law phase, we investigate whether our
CSM model affects the behaviour of the light curve during the Lira law
phase. We find that LEs make the opposite, they tend to smooth the
color evolution. This goes in the opposite direction of our goal of
finding light echoes in fast Lira decliners spectra and favors the DS
model.

\begin{figure}[h!]
\centering
\includegraphics[width=1.0\linewidth]{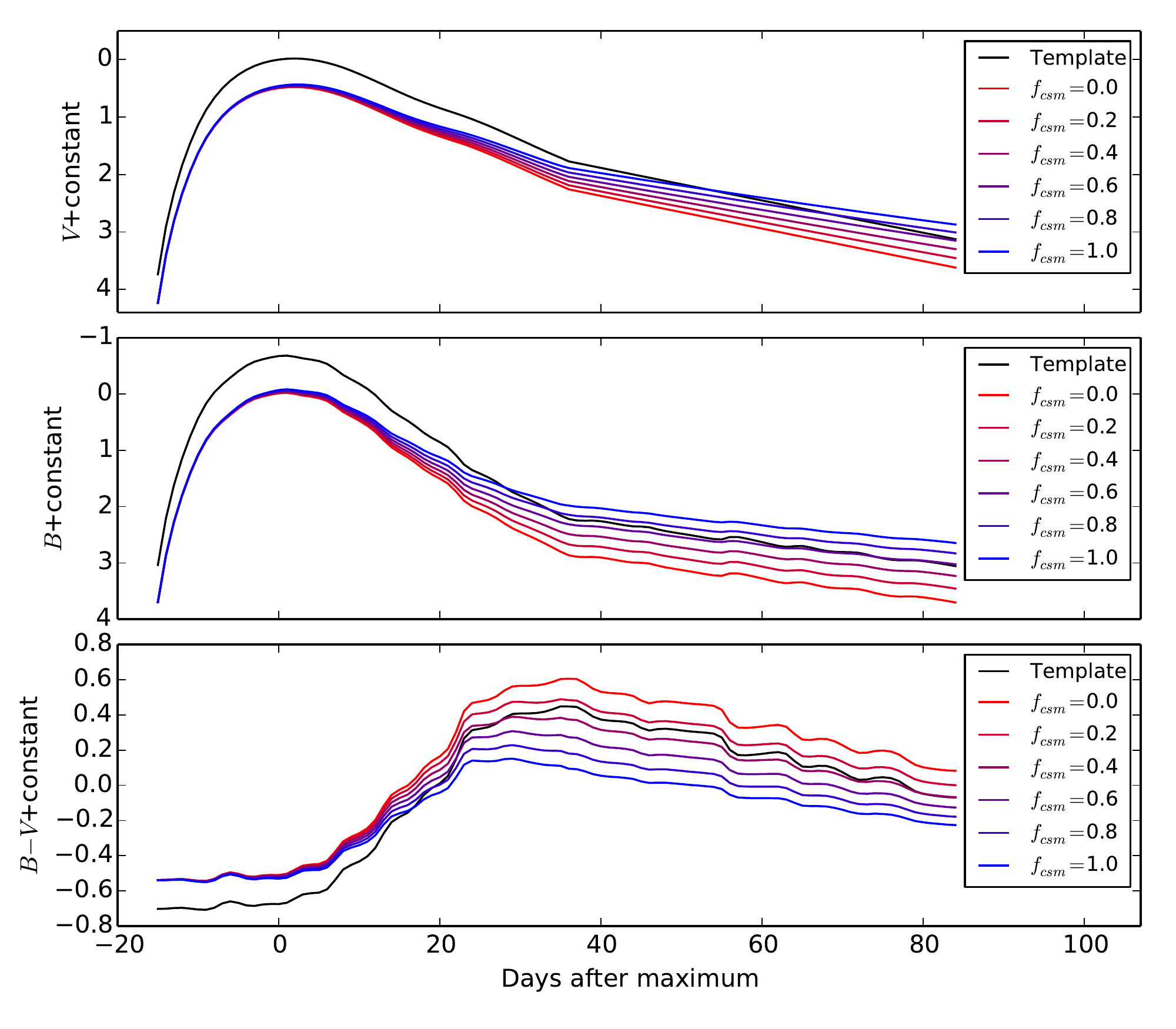}
\caption{Simulated light curves using our LE model. In black is the
  original light curve without extinction. The reddest curve represents
  a light curve with pure extinction ($A_V$=0.5 and $R_V$=3.1) and the
  rest are scenarios with different fractions of
  CSM.}\label{fig:lightcurves}
\end{figure} 

In Figure \ref{fig:lightcurves} we present these simulated light
curves. As expected, in the pure extinction scenario the $B$ and $V$
light curves (red line) are just uniformly shifted downward with
respect to the original light curve without extinction (black
line). On the other hand, the presence of LEs due to CSM modify the
shape of these curves, increasing the brightness in $B$ and $V$ at
later epochs as blue light from maximum light is reaching the
observer. However, the slope of the $B-V$ evolution actually becomes
shallower during the Lira law, see Table \ref{table:slopes}. This can
be explained qualitatively as our LE model adds light emitted at
previous epochs to the intrinsic emitted light, including late time
emission with small time delays. These contributions make the observed
light curve evolve slowly.

The wiggles in the $B$ and $B-V$ light curves are not real. They are
caused by the way we compute the $B$ magnitudes from the template
spectra, which depend on the available SN spectra at each time
window. On the other hand, the $V$ magnitudes match the observed
photometry by construction. Despite this, the general shape of the
light curves is clear.

\begin{table}[ht]
\caption{$B-V$ slope during the Lira law phase from simulated light curves.}
\centering
\label{table:slopes}
\begin{tabular}{clc}
\hline
\hline
 $f_{\mathrm{CSM}}$  & Slope $\pm$ error [mag/day]  \\
\hline
0.0 & $-0.010\pm 0.002$ \\
0.2 & $-0.008\pm 0.001$ \\
0.4 & $-0.008\pm 0.001$ \\
0.6 & $-0.007\pm 0.001$ \\
0.8 & $-0.004\pm 0.001$ \\
1.0 & $-0.005\pm 0.001$ \\
\hline 
\end{tabular}
\end{table}

\subsection{Dust sublimation and its effects on the light curve}

\begin{figure}[h!]
\centering
\includegraphics[width=1.0\linewidth]{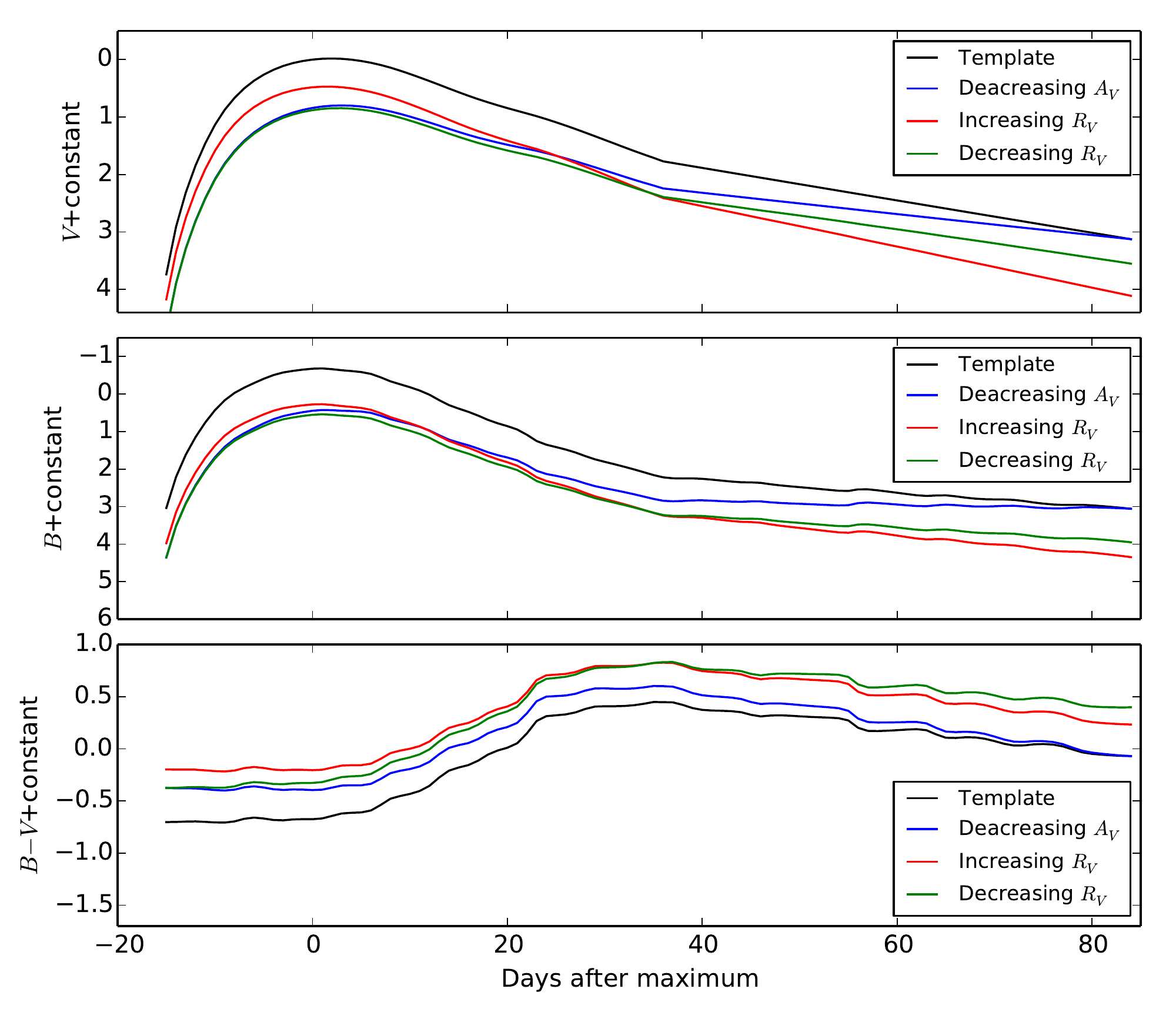}
\caption{Simulated light curves using extinction laws that vary over
  time. In black is the original light curve without extinction. The
  blue curve is a model with Av decreasing from 0.5 to 0.0 and
  $R_V=3.1$ held constant. The red curve represents a light curve with
  $A_V$=0.5 and $R_V$ evolving from 2.0 to 3.1. The green curve
  represents a light curve with $A_V$=0.5 and $R_V$ evolving from 3.1
  to 2.0.}\label{fig:lightcurves_ext}
\end{figure}

A decreasing extinction or opacity could occur if the CSM dust that
was extincted at maximum light got sublimated by an increment of
temperature due to the SN radiation, reducing the total opacity in the
line of sight. This sublimation could also change the observed $R_{V}$
as it might change the grain size distribution and composition,
explaining the evolution found in F13. If this happens during the Lira
law phase, it will be reflected in a steeper $B-V$ slope. To test this
hypothesis we simulated light curves with a decreasing $A_V$, and
variable $R_V$. Figure \ref{fig:lightcurves_ext} presents these three
different scenarios. We found that a decreasing $A_V$ or increasing
$R_V$ can make the $B-V$ evolution become steeper.

A smaller $R_V$ makes the extinction law more sensitive to the blue
than the red, which is expected if the grain size distribution favors
small sizes. An increasing $R_V$ could be produced if the smaller
grains disappear as sublimation occurs. Another possibility is that
the intrinsic CSM $R_V$ was lower than the ISM $R_V$, therefore as the
circumstellar dust is being sublimated, the total $R_V$ increases
reaching values similar to the interstellar $R_V$. This hypothesis is
also consistent with the mean $\Delta A_{V}$ found at different epochs
for fast Lira decliners.

Within this context, it is important to note that recently
\citet{Goobar2014b} showed that SN2014J, an extincted red SN Ia
with low $R_V$ \citep{Foley2014} and with strong narrow absorption
features \citep{Welty2014}, has possible signatures of cooling from
shocked material from nearby CSM of dimensions larger than 1
$R_\odot$. But the scales of this CSM are smaller than the ones
considered in this paper.

Another possibility is that the radiation pressure (RP) of the SN is
blowing away the CSM dust particles. If the CSM radius is increased,
the observed extinction will decrease as the column density decreases,
also producing blueshifted absorption lines. A rough estimation of the
time scales in which the RP could produce an observable change in the
CSM extinction can be calculated as
\begin{eqnarray}
\tau_{\mathrm{RP}} &=& \frac{R_{\mathrm{CSM}}}{\Delta v} \\
\tau_{\mathrm{RP}} &=& \frac{m_{d} \ R_{\mathrm{CSM}} \ c}{f_{\mathrm{SN}} \ \sigma_a \ \Delta t}
\end{eqnarray}
where $R$ is CSM radius, $\Delta v$ is a characteristic velocity of
the dust grains after they absorb linear momentum from the SN
incident radiation flux $f_{\mathrm{SN}}$ that can be calculated using
the well known relation between the radiation pressure and flux of an
electromagnetic wave. We consider spherical dust particles of radius
$a$, internal density of 1 gr/cm$^3$, mass $m_{d}$, and with a cross
section $\sigma_a$ calculated using the Mie theory.  Assuming a
typical SN Ia luminosity and a time range $\Delta t$= 25 days,
centered at maximum light, in which the RP injects momentum to the
dust particles, we can obtain an estimate of the time scale in which
this effect could be observed:

\begin{equation}
\tau_{\mathrm{RP}}=1.4\times 10^2 \left(\frac{a}{1 \mu m}\right)\left(\frac{R}{0.01 \mathrm{pc}} \right)  \mathrm{days}
\end{equation}

Thus, if the radiation pressure is the cause of the decreasing
extinction at $\tau_{\mathrm{RP}} \sim$ 80 days past explosion (Lira
law), the CSM dust particles should be smaller than 1 $\mu$m or be at
distances of $\sim10^3$ astronomical units. At these distances we
expect that the sublimation time scales of these smaller grains are
much shorter than the calculated above.

Therefore, even if dust sublimation time scales are too long to
account for the changes in extinction and $R_V$, during the Lira law
phase, we expect that the CSM expansion due to RP should increase the
observed $R_V$ and decrease the extinction, as smaller grains are
blown away.

An alternative scenario to the CSM sublimation and RP that could
explain the decrease in extinction is the transversal expansion of the
ejecta in a non-homogeneous ISM. This possibility is explored in F13
where they conclude that this scenario could explain the change in the
average column density as the photospheric radius increases, but it
does not explain the change in the observed $R_{V}$ over time.




\section{Conclusions} \label{conclusions}

A CSM model producing light echoes (LEs) has been developed. It is
simple enough to be computed quickly and be used in our fitting
routines, but with enough complexity to account for the albedo of the
dust and the scattering phase function. Our model predicts two
spectral signatures produced by LEs at 4100 and 6200 \AA \ that can
help us discriminate between a pure extinction or extinction+LEs
scenario. These features appear within small wavelength ranges as
opposed to overall color changes that can also be produced by
reddening. The evolution of these signatures is another tracer of the
presence of CSM producing LEs.

We compare our models with observed SN spectra. We find that LEs from
CSM at 0.01-0.25pc are not a global phenomenon in fast Lira decliner
SNe when they are compared to slow Lira decliners. ISM or CSM dust
being sublimated (DS) at later epochs explains better the observed
spectra when the models are fitted using the overall
spectrum. Additionally, we find no evidence for LEs based on the
narrow spectral diagnostics predicted by our model.

We explore the effects on the light curve of circumstellar dust being
sublimated or blown away by radiation pressure, finding that both
scenarios could produce a faster $B-V$ decline during the Lira law and
a change in $R_V$, although a more rigorous physical modeling is
needed to explore these possibilities.

We laid out several ways to improve our models: adding a $R_{V}$ as a
free parameter in our fits; using different CSM geometries to test our
predictions; and a Monte Carlo radiative transfer simulation to see if
the LE signatures remain in the multiple scattering scenario. The
analysis can also be improved using a larger sample of SNe,
particularly those highly extincted and with possible CSM
characteristics.

\label{con}

\section*{Acknowledgement}

We thank the referee for providing constructive comments and help in
improving the contents of this paper. We also thank K. Maeda for useful
discussion. SM would like to thank to
CONICYT-PCHA/MagísterNacional/2014 - folio 22140628. SG thanks CONICYT
through FONDECYT grant 3130680 for its support. FF thanks CONICYT
through FONDECYT grant 11130228 for its support. SM, SG, FF and MH
acknowledge support provided by the Ministry of Economy, Development,
and Tourism's Millennium Science Initiative through grant IC120009,
awarded to The Millennium Institute of Astrophysics,
MAS. E.~Y.~H. acknowledges the generous support provided by the Danish
Agency for Science and Technology and Innovation through a Sapere Aude
Level 2 grant.

\bibliographystyle{apj}
\bibliography{astro}

\appendix

\section{Average spectra per SN per time range} \label{A:average_spectra}
For the $jth$ spectrum of the $ith$ SN we defined the following weights:
\begin{equation}
w_{j}^{i}=e^{-(t_{j}-t_{0})^{2}/2\sigma_{t}^{2}}
\end{equation}
where $t_{0}$ is the epoch that we want to represent and $\sigma_{t}=5.0$ days is half of the time window. The average spectrum for a particular SN at the given epoch $t_0$ was then calculated according to 
\begin{equation}
f^{i}(\lambda)=\frac{\sum_{j}w_{j}^{i}f_{j}^{i}(\lambda)/\delta f_{j}^{i}(\lambda)^{2}}{\sum_{j}w_{j}^{i}/\delta f_{j}^{i}(\lambda)^{2}} \label{eq:average_spec}
\end{equation}
where $f_{j}^{i}(\lambda)$ and $\delta f_{j}^{i}(\lambda)$ are the normalized flux of the $jth$ spectrum and its error. The sum goes over all the available spectra of the $ith$ SN with measured fluxes in the given time range. We also computed an error and a representative epoch for each average spectrum in a similar way.

\section{Template spectra} \label{A:template}
To construct template spectra we define two weight factors for time and stretch of the $ith$ average spectrum of each slow Lira decliner SN:  
\begin{eqnarray}
w_{t}^{i}&=&e^{-(t_{i}-t_{0})^{2}/2\sigma_{t}^{2}}\\
w_{s}^{i}&=&e^{-(s_{i}-s_{0})^{2}/2\sigma_{s}^{2}}
\end{eqnarray} 
We chose $\sigma_{t}=1.5$ days and $\sigma_{s}=0.11$ in order to reproduce a specific epoch and stretch. We do not have strong arguments to choose particular values, thus we use the standard deviation of stretchs in our sample for $\sigma_{s}$ and a $\sigma_{t}$ value smaller than our time windows, but large enough to reproduce smooth light curves. Defining  $\alpha_{i}=w_{t}^{i}\times w_{s}^{i}$ the template spectrum with a certain epoch $t_{0}$ and stretch $s_{0}$ is
\begin{equation}
F(\lambda)=\frac{\sum_{i}\alpha_{i}f_{i}(\lambda)/\delta f_{i}(\lambda)^{2}}{\sum_{i}\alpha_{i}/\delta f_{i}(\lambda)^{2}}
\end{equation}
where the sum goes over all the available already averaged spectra from slow Lira decliners with measured fluxes at wavelength $\lambda$. Finally we normalized the template spectrum by its flux in the $V$ band.

\begin{figure}[h!]
\centering
\includegraphics[width=0.5 \linewidth]{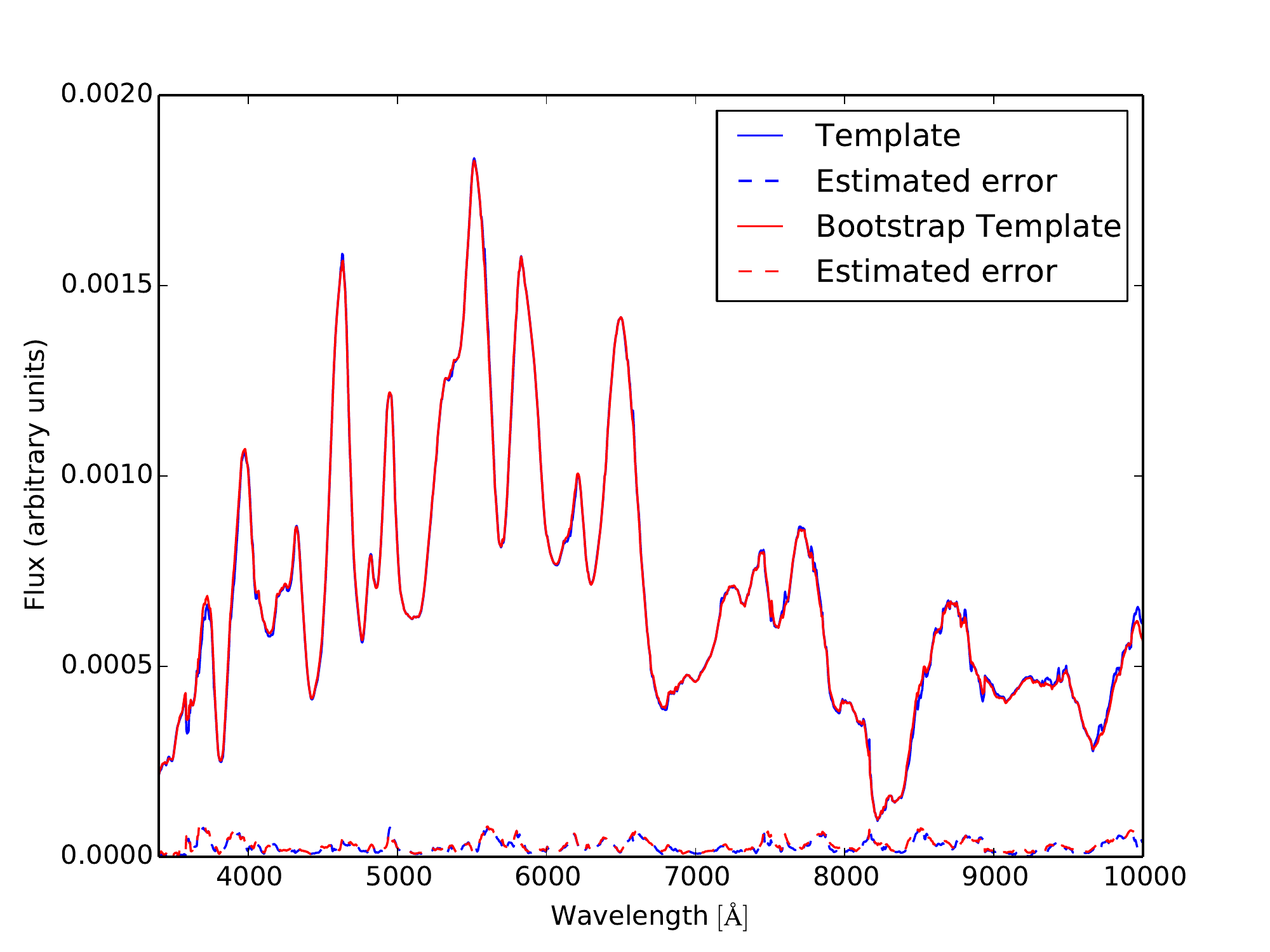}
\caption{Template spectrum of a slow Lira decliner SN using weighted averages vs using the bootstrap technique at 50 days after maximum light and using a stretch of 0.98.}\label{bootstemp}
\end{figure}

To ensure that these average templates are not heavily biased by few extreme SNe, we performed a bootstrap simulation. Then computing the mean template and the dispersion around it, we obtained a ``bootstrap template'' and its error. In figure \ref{bootstemp} we compare a template constructed using weighted averages and a template using the bootstrap simulation. In both cases we used all the slow Lira decliners sample. There are slight differences between both spectra, but these are not significant and the estimated errors are very similar. 

\section{Extinction law fit} \label{A:ext}
\begin{figure}[h]
\centering
\includegraphics[width=0.5\linewidth]{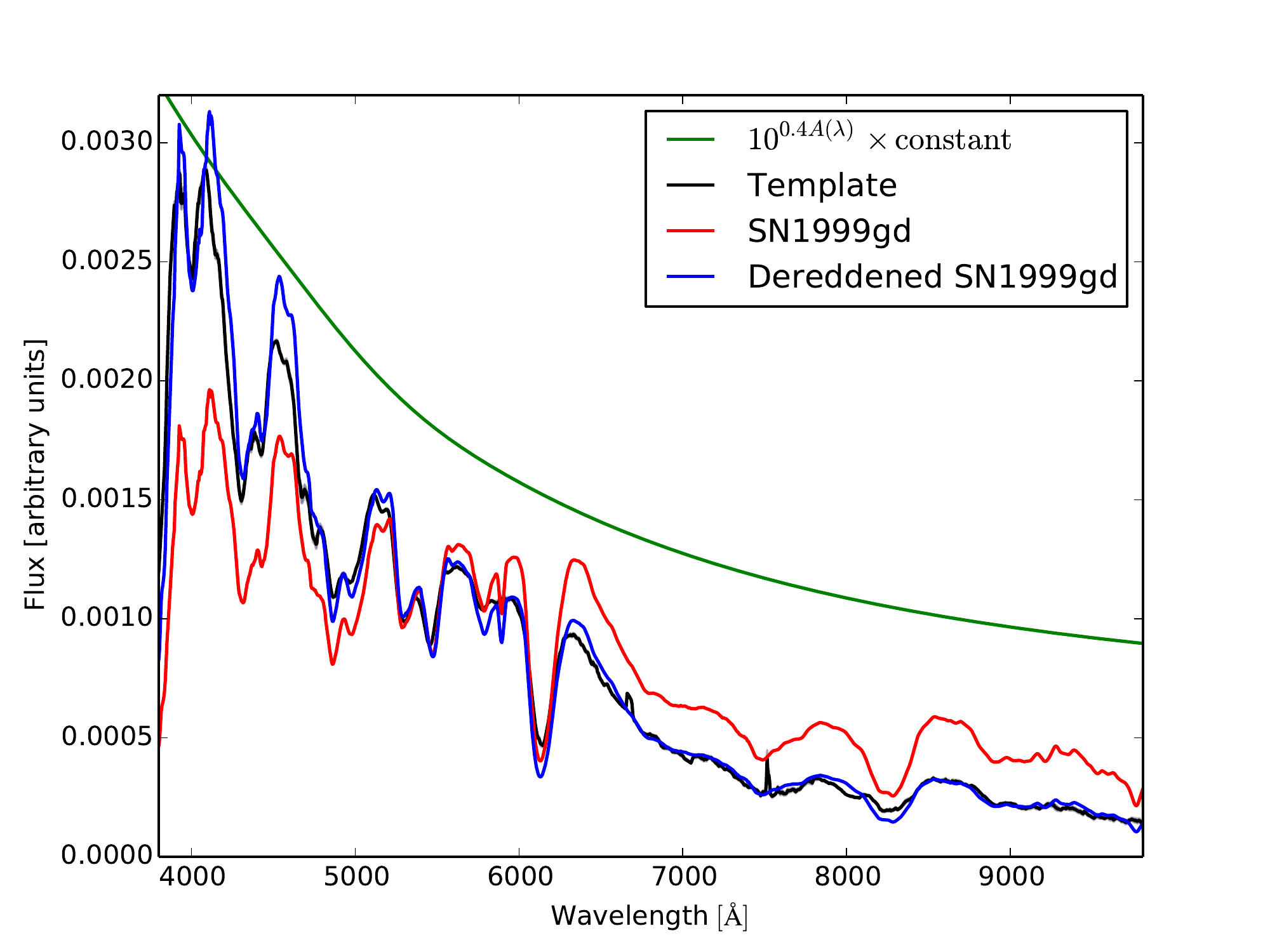}
\caption{Extinction law fit for SN1999gd at maximum light using \citet{Fitzpatrick99} extinction law. The best fit with $R_V=3.1$ yields $A_V=1.280\pm 0.003$. In red we present the observed spectrum of SN1999gd at maximum light, in black a constructed template representing the same epoch and stretch. The blue line is the dereddened spectrum and in green, the extinction factor $e^{-\tau_{\lambda}}$ or $10^{-0.4A_{\lambda}}$ shape.\label{ejemext}}
\end{figure}

To fit an extinction law we minimize a chi-square function for each average spectrum $i$
\begin{equation}
\chi^{2}_i=\sum_{\lambda}\frac{(f^{i}(\lambda)-f^0(\lambda)10^{-0.4A(\lambda)}/C_{1})^{2}}{\delta f^{i}(\lambda)^{2}+(\frac{f^{i}(\lambda)}{f^0(\lambda)}\delta f^0(\lambda))^{2}} \label{eq:chi_ext}
\end{equation}
where $f^{i}(\lambda)$ is the normalized flux of the $ith$ fast Lira decliner SN at maximum light and $f^0(\lambda)$ is an unreddened template representing the same intrinsic flux. In figure \ref{ejemext} we present a particular extinction law fit for SN1999gd at maximum light. We found that $A_{V}=1.280 \pm 0.003$ best represents the extinction using our template spectrum. 

\section{Light echo fit} \label{A:LEfit}
To fit our light echo model we introduce a new factor $N$ to force the spectra to evolve consistently with the light curve in the $V$ band. Hence we calculate $S(t,\lambda)$ as
\begin{eqnarray}
\mbox{\small $S(t)=\frac{w(1-10^{-0.4Af_{\mathrm{CSM}}})}{\tau_{\max}} \int_{0}^{\tau_{\max}} N(t,\tau) f^{0}(t-\tau)\Phi'(\tau) d\tau \hspace{0.5cm} $} \\
\mbox{\small $N(t,\tau)=10^{-0.4(V(t-\tau)-V(t))} $}  \hspace{4cm}
\end{eqnarray}
where we omitted the wavelength dependency of $S$, $f^0$, $\omega$, $\Phi'$ and $A$ (the extinction law found at maximum light). $f^{0}(t)$ is a template spectrum representing the same intrinsic flux at time $t$. We multiply each spectrum by $N(t,tr)$, which is the $V$ light curve normalized in $t$, to correct for the fact that all our templates are normalized by their flux in the $V$ band. This factor is computed using weighted averages of the $V$ magnitudes of slow Lira decliners considering epoch and stretch as we did for the template spectra. The $V$ magnitude data was taken from SiFTO fits to the data \citep{Conley08}.

\end{document}